\documentclass[11pt]{article}
\usepackage{amsmath}
\usepackage{amssymb}
\usepackage{graphics}
\begin{document}

\title{Metastability of $R$-Charged Black Holes}
\author{D.~Yamada
        \bigskip
        \\
       {\it Racah Institute of Physics},
       {\it The Hebrew University of Jerusalem},
        \smallskip
       \\
         {\it Givat Ram, Jerusalem, 91904 Israel }
         ({\tt daisuke@phys.huji.ac.il})}
\date{}
\maketitle

\begin{abstract}
The global stability of $R$-charged AdS black holes 
in a grand canonical
ensemble is examined by 
eliminating the constraints from the action,
but without solving the equations of motion,
thereby constructing the reduced action of the system.
The metastability of the system is found to set in
at a critical value of the chemical potential which is
conjugate to the $R$-charge.
The relation among the small black hole, large black hole
and the instability is discussed.
The result is consistent with the metastability found
in the AdS/CFT-conjectured dual field theory.
The ``renormalized'' temperature of AdS black holes, 
which has been rather \textit{ad hoc},
is suggested to be the boundary temperature in the sense of
AdS/CFT correspondence.
As a byproduct of the analysis,
we find a more general solution
of the theory and its properties are briefly discussed.

\end{abstract}


\tableofcontents

\section{Introduction}
The global structure of a black hole action is the interest
of this paper.
Examination of
this, in general, is a formidable task and the difficulties include
the large number of the fields involved and the complicated nature
of the action itself.
For instance, in a single
scalar theory
at classical level, one simply plots the potential of the
theory with respect to the expectation value of the scalar field
and examines the global minimum of the theory.
Similar program for a theory that involves gravity is not
as straightforward due to the complications.
Therefore, a black hole action is commonly studied by evaluating
the values of it on known black hole solutions, by this means
constructing the on-shell action of the theory.
The on-shell action is valid in an infinitesimal neighborhood
of the black hole saddle point
and the local behavior of black holes, such as the first law
of thermodynamics and local thermodynamic stability, can be studied
from the action.
This method, however, is not adequate for examining the global
structure of a black hole action.

In Reference~\cite{Braden:1990hw}, Braden {\it et al.} devised
the technique that yields a simplified off-shell action,
called ``reduced action'' of a black hole system.
The procedure for constructing the reduced action is as follows.
One starts with the action of a system and place it in a finite
box.
Then, restrict the analysis to a class of geometries
that satisfy certain conditions, such as static spherical symmetry.
The restriction results in a specific form of the metric and
this is analogous to a metric ansatz that one makes in solving
the Einstein equation.
In addition to the geometrical restriction, one imposes the black
hole ``cigar'' topology on the spacetime manifold, giving
a condition that the metric must satisfy.
Next, the thermodynamic data is encoded on the wall of the box.
The data, in the case of a grand canonical ensemble, is the
temperature and chemical potential.
(We will show exactly how to encode the data in
Section~\ref{sec:boundaryData}).
As will be explained in Section~\ref{sec:constraints}, the
physical states of the
system in general obeys constraint equations
including the Hamiltonian constraint, and
they are solved and eliminated from the action.
Finally, the action functional is integrated 
with the topology condition
and the boundary data, resulting in the simplified
reduced action.
Notice that the equations of motion have not been used in reducing
the action.
Thus the reduced action is valid throughout
the physical
parameter space and a black hole solution corresponds to
a saddle point of the action in the parameter space.
Therefore, this action
captures the global structure of the theory.

After reviewing the five dimensional $R$-charged black hole
solution and its properties in Section~\ref{sec:review},
in Section~\ref{sec:construction},
we apply the procedure sketched in the previous paragraph
to this black hole system in a grand canonical
ensemble.
This black hole system can be obtained through $S^5$
compactification of ten dimensional Type IIB supergravity
with gauged $SO(6)$ global symmetry
\cite{Pernici:1985ju,Gunaydin:1984qu}.
A special case with decoupled scalars can be obtained rather
simply by
``twisting'' the five sphere upon the compactification,
as demonstrated by Chamblin {\it et al.} in
Reference~\cite{Chamblin:1999tk}.
It has also been shown by Cveti\v c and Gubser
in Reference~\cite{Cvetic:1999ne} that
the black hole geometry arises from the
near horizon limit of rotating D3-branes in ten dimensions
with the rotation in their transverse directions.
Therefore this system has drawn numerous interests
in connection to string theory
and the examination of the global structure is worthwhile
in its own right.
However, there is an additional and more specific
motivation for studying the global stability of the theory.

According to the AdS/CFT correspondence
\cite{Maldacena:1997re,Gubser:1998bc,Witten:1998qj},
the $R$-charged black hole system in a grand canonical ensemble
is conjectured to be the dual of
strongly coupled thermal $\mathcal{N}=4$ supersymmetric Yang-Mills
theory in the large $N_c$ limit
with the chemical potentials corresponding to the maximal
Abelian subgroup
of the $SU(4)$ $R$-summery. 
The weak coupling analysis of the field theory in
Reference~\cite{Yamada:2006rx} shows that, despite the difference
in the coupling regime,
the phase diagram in the parameter space of temperature and 
chemical potential closely
resembles the one obtained in the gravity analysis of
Reference~\cite{Cvetic:1999ne}.
However, in the phase diagram of the field theory,
the additional structure has been found where the equilibrium state
of the system becomes metastable beyond a certain critical value
of the chemical potential.

The metastability in the field theory
was found by examining the global behavior of the
scalar field effective potential at one-loop level.
When the system is in the deconfined plasma phase with
a sufficiently low value of the chemical potential,
the origin of the scalar field space is the absolute
minimum of the (one-loop) effective potential and defines the
stable ground state of the theory.
When the chemical potential is raised beyond the critical value,
the effective potential goes down to minus infinity at the
asymptotic region of the scalar expectation value.
The minimum at the origin persists to exist in this phase but
it now only is a local minimum with a potential barrier
separating it from the unstable direction.
The state at the origin, therefore, is metastable in this phase
and the true ground state of the theory does not exist.
Further rise in the chemical potential would eventually lead
to the local instability of the state where the barrier that
separates the local minimum and the unstable direction disappears.

We suspect that the metastability is caused by the violation
of the BPS inequality expected for the states of
the field theory.
The chiral primary fields of $\mathcal{N}=4$ super-Yang-Mills
theory contain the half-BPS operators that are characterized
by their scaling dimension equaling their $R$-charges and
in general, the dimension of an operator must be larger than
or equal to the charge.
When the theory is conformally mapped to a sphere, the operator
and the dimension are mapped to a state of the theory and
the energy of the state, respectively.
Therefore, the half-BPS states have the energy being equal
to the $R$-charges.
When we introduce the chemical potential conjugate to the
charge, we expect that the product of the charge and the chemical
potential contributes to the total energy of the state.
(Such an argument has been made in Reference~\cite{Yamada:2005um}
and further generalized in \cite{Harmark:2006di}.)
Hence if the chemical potential is raised higher than one (in
the units of the radius of the sphere),
we expect that the BPS inequality is violated
and something should go wrong with the theory.
The metastability of the field theory analysis sets in exactly
at this critical value of the chemical potential.

Since the dimension of the BPS operator is protected against
the quantum corrections and change in the coupling constant,
we expect
similar behavior in the gravitational analysis which is
supposed to be the strong coupling region of the same theory.
But it is clear that the local on-shell 
analysis that has been carried
out in Reference~\cite{Cvetic:1999ne} is not able to detect the
metastability of the black hole system.
This is why we embark on the program outlined above.
Analyzing the resulting reduced action, we will find the metastability
in Section~\ref{sec:metastability}.

In Section~\ref{sec:discussions}, a few issues that come up during
the analysis are discussed.
First in Section~\ref{subsec:adsCftBoundaryTemp}, the mysterious
``renormalized'' temperature that has been adopted in literature
on the AdS black hole systems is suggested to be naturally understood
in the context of the AdS/CFT correspondence.
Then Section~\ref{subsec:moreGeneralSolution} presents a more general
solution of the theory than what has been known.
This solution is obtained as a byproduct of the main analysis and
its properties are briefly discussed.
Section~\ref{subsec:finiteCounterterm} addresses a subtle issue
of the physical interpretation of 
the counterterm that does not
diverge as the wall of the box is pushed to infinity.
Finally in Section~\ref{subsec:comparison}, we contrast the
revised phase diagram of the gravity side to the one in the
field theory side.

\section{Review of $R$-Charged Black Holes}\label{sec:review}
We start by reviewing the $R$-charged black hole solution and their
thermodynamics,
mainly following Behrndt {\it et al.} \cite{Behrndt:1998jd} and
Cveti\v c and Gubser \cite{Cvetic:1999ne}.
The analysis on the instability of the black holes
is slightly extended from the work of Cveti\v c and Gubser
\cite{Cvetic:1999ne} for the cases with different charge 
configurations.%
\footnote{
  The extended results were also reported in \cite{Yamada:2006rx}.
}

The $R$-charged black hole of our interest is the solution to
five dimensional $\mathcal{N}=8$ gauged supergravity
where the gauged symmetry is the $U(1)^3$ truncation of
the $SO(6)$ global symmetry that arises from the $S^5$
compactification of ten dimensional supergravity.
The solution is given by the metric
\begin{equation}\label{eq:metricSolution}
  ds^2 = - H(r)^{-2/3} f(r) dt^2
        + H(r)^{1/3} \big[ f(r)^{-1}dr^2 + r^2 d\Omega_3^2 \big]
        \;,
\end{equation}
where $d\Omega_3^2$ is a metric on the unit three sphere and
the function $H(r)$ is
\begin{equation}
  H(r) = H_1(r)H_2(r)H_3(r)
          \quad
  \text{with}
          \quad
  H_j(r) = 1 + \frac{q_j}{r^2}
        \;,
\end{equation}
and
\begin{equation}
  f(r) = 1 - \frac{r_0^2}{r^2} + \frac{r^2}{R^2}H(r)
        \;.
\end{equation}
The physical interpretations of the parameters 
$q_j$, $r_0$ and $R$
are as follows.
As mentioned in the introduction,
it is possible to interpret this solution as the near horizon
limit of rotating D3-branes in ten dimensions and in this
perspective,
the parameters $q_j$ with the dimension length squared
are related to the three possible
transverse angular momenta of the D3-branes.
(See Reference \cite{Cvetic:1999ne}.)
The parameter $r_0$ in the function
$f$ measures the ``non-extremality''
of the black hole in that $r_0 = 0$ corresponds to the
extremal BPS-saturated solution. The non-extremality parameter
can be expressed in terms of the horizon radius $r_H$ by
solving the equation $f(r_H) = 0$, explicitly,
\begin{equation}\label{eq:m}
  r_0^2 = r_H^2 \bigg\{ 1 + \frac{r_H^2}{R^2}H(r_H) \bigg\}
        \;.
\end{equation}
The parameter $R$ is the length scale of the theory.
Note that the solution approaches to the AdS metric
at the $r\to\infty$ asymptotic region and we see
that the parameter $R$ is the curvature radius of
this asymptotic AdS space.
Therefore, it is also related to the cosmological constant.

This theory contains scalar fields, $X_j(r)$ with $j=1,\cdots ,3$,
satisfying the constraint%
\footnote{
  As in Reference~\cite{Behrndt:1998jd} the location of the
  index $j$, either up or down, {\it does} matter sometimes.
  In this work we exclusively use the lower index, {\it which
  corresponds to the upper index of the reference}, purely for
  the presentation reason.
  The notation is consistent within this work, however, if
  the reader is interested in comparing expressions with the
  reference, the convention difference must be watched.
}
\begin{equation}
  X_1(r)X_2(r)X_3(r) \equiv 1
        \;.
\end{equation}
The solutions to the scalar fields are
\begin{equation}
  X_j(r) = H_j(r)^{-1}H(r)^{1/3}
        \;.
\end{equation}

This theory contains three dimensionless Abelian gauge
fields arising from the gauged
$U(1)^3$ and the solution takes the form
\begin{equation}\label{eq:lorentzianA}
  A_{j\,\nu}(r) = \delta_{\nu,0}
        \bigg( \frac{e_j}{r_H^2 + q_j} - \frac{e_j}{r^2 + q_j} \bigg)
        \;,
\end{equation}
where $\nu$ is a spacetime index, the parameters
$e_j$ with the dimension length squared are the electric
charges of the system under the Abelian fields,
the index $j$ runs, again, from $1$ to $3$ and
we have set $A_{j\,\nu}(r_H)=0$ for
the reference point of the potentials.
Observe that the time-component of the
gauge fields at the coordinate $r$
is the electric voltage
between the horizon and the location $r$.
We denote the potentials at the boundary
with $r=\infty$
by $\mu_j$ and they are
\begin{equation}\label{eq:chemp}
  \mu_j = \frac{1}{R}\,\frac{e_j}{r_H^2 + q_j}
        \;,
\end{equation}
where the length scale $R$ is inserted for the dimensional reason.
The parameters $q_j$ and $e_j$ are related through
\begin{equation}\label{eq:eqRelations}
  e_j^2 = q_j (r_H^2 + q_j)
        \bigg[ 1 + \frac{r_H^2}{R^2}\prod_{k\neq j}H_k(r_H) \bigg]
        \;,
\end{equation}
where the repeated index $j$ is not summed over.

The ADM mass $M$ of the black hole is
\begin{equation}\label{eq:MinI}
  M = \frac{\omega_3}{8\pi G_5}
        \bigg( \frac{3}{2}r_0^2 + \sum_{j=1}^3 q_j \bigg)
        \;,
\end{equation}
where $\omega_3 = 2\pi^2$ is the volume of the unit three sphere
and $G_5$ is the five dimensional Newton's constant which has
the dimension of length cubed.
This ADM mass assumes the AdS background without the presence of
the black hole, {\it i.e.}, the mass $M$ is the mass difference
between the AdS black hole spacetime and the thermal AdS space without
the black hole.%
\footnote{
  In Reference \cite{Buchel:2003re}, Buchel and Pando Zayas point
  out that the ADM mass actually is divergent due to the presence of
  the gauge fields.
  However, the Brown-York \cite{Brown:1992br}
  type computation by Liu and Sabra  \cite{Liu:2004it}
  who include the ``finite counterterm''
  yields the mass as presented in Equation~(\ref{eq:MinI}).
  The finite counterterm is discussed further in
  Section\ref{subsec:finiteCounterterm}.
}
The entropy $S$ is a quarter of the horizon area as usual:
\begin{equation}
  S = \frac{A}{4G_5} = \frac{2\pi^2r_H^3\sqrt{H(r_H)}}{4G_5}
        \;.
\end{equation}
The inverse Hawking temperature $\beta_H$ is given by the expression
\begin{equation}\label{eq:invTemp}
  \beta_H = \frac{2\pi R^2r_H^5\sqrt{H(r_H)}}
                {2r_H^6 + r_H^4 (R^2 + \sum_{j=1}^3q_j) - \prod_{j=1}^3q_j}
        \;.
\end{equation}

In Reference \cite{Cvetic:1999ne}, the thermodynamic properties of
the black hole in the grand canonical ensemble is examined by
introducing the Gibbs free energy
\begin{equation}\label{eq:Gibbs}
  I_{\text{Gibbs}} = \beta_H \bigg( M - \sum_{j=1}^3 e_j\mu_j \bigg)
                       - S
        \;,
\end{equation}
where the length scale $R$ is set to $1$.
The parameters
$\beta_H$ and $\mu_j$ are the fixed data of the ensemble
and the black hole corresponds to the saddle point of this
function.
Notice that the potentials $\mu_j$ are taken to be
the chemical potentials of the grand canonical ensemble.

As mentioned earlier, the ADM mass is the difference between
the black hole spacetime and the thermal AdS space without black holes.
Thus, a change in the sign of the Gibbs free energy indicates
that the
ground state (or thermal equilibrium state) of the system switches
from the black hole to the thermal AdS or {\it vice versa}.
This is the well-known Hawking-Page phase transition \cite{Hawking:1982dh}.
The phase transition line of the phase diagram in $(\mu,T)$
parameter space can therefore be determined
by the condition $I_{\text{Gibbs}}=0$.

Cveti\v c and Gubser \cite{Cvetic:1999ne}
further examined the local thermodynamic stability of the
black hole solutions.
At an instability point, the black hole solution that extremizes
the Gibbs free energy becomes an unstable saddle point, as opposed to 
a local minimum.
Therefore, when the Hessian of $I_{\text{Gibbs}}$ with respect to $r_H$ and $q_j$
is evaluated at the black hole solution,
the determinant of the Hessian vanishes at the critical 
local instability point.
This condition establishes the instability line in the phase diagram.
Even though this criterion does not determine which side of the
instability line corresponds to the stable black hole,
one can figure that by knowing that the region of high temperature
and zero chemical potential must correspond to a stable black
hole solution.

\begin{figure}[h]
{
\centerline{\scalebox{0.45}{\includegraphics{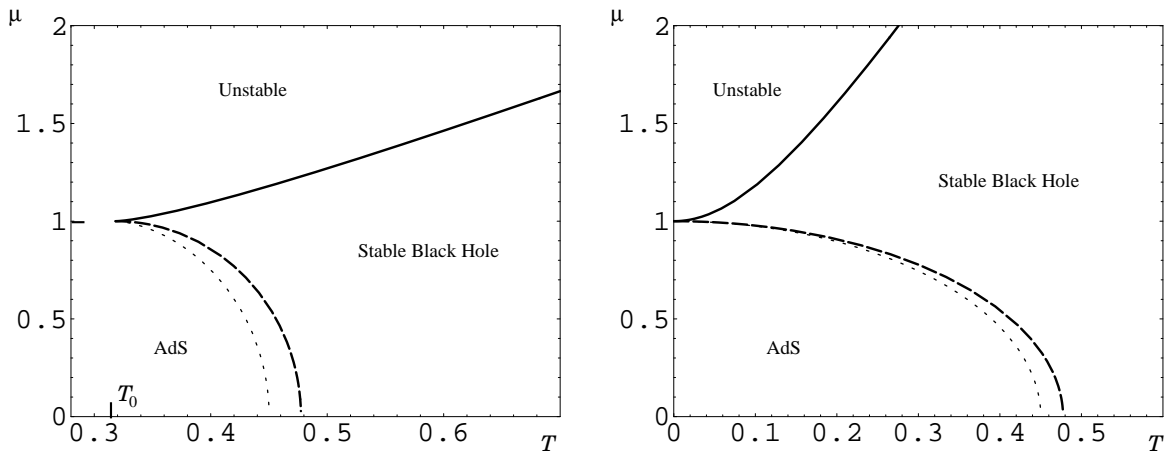}}}
\caption{\footnotesize
  The phase diagram on the left is plotted for
  $(q_1,q_2,q_3)=(q,0,0)$ and on the right is for $(q,q,q)$.
  Both $\mu$ and $T$ are measured in the units of $1/R$,
  the inverse of the AdS curvature radius.
  The solid and dotted lines are the black hole instability
  lines whereas the dashed lines are the Hawking-Page phase
  transition lines.
  The regions between the dashed and dotted lines indicate the
  spinodal phases where the AdS space is energetically
  preferred but the black hole still remains locally stable.
  The temperature $T_0 = 1/\pi$ on the left-hand diagram is
  where $r_H=0$.
}\label{fig:holethermo}%
}
\end{figure}
The resulting phase diagrams are plotted in Figure~\ref{fig:holethermo}.
We choose to plot the phase diagram in $\mu$-$T$ plane as
oppose to Reference \cite{Cvetic:1999ne} whose choice is
$q$-$r_H$ plane.
We have also extended the analysis of the same reference to
the cases with more than one $q$ is turned on.
The left-hand diagram is plotted for
$(q_1,q_2,q_3)=(q,0,0)$ and the right is for $(q,q,q)$.
The solid and dotted lines are the black hole instability
lines whereas the dashed lines are the Hawking-Page phase
transition lines.
The regions between the dashed and dotted lines indicate the
spinodal phases where the AdS space is energetically
preferred but the black hole still remains locally stable.
The general structure of the phase diagrams are similar
for the different charge configurations
[including the case with $(q,q,0)$].
However the low temperature regions are quite different.
The temperature $T_0$ in the left-hand diagram denotes the Hawking
temperature at which $r_H=0$.
When we have three equal non-zero charges, $T_0 = 0$, but
for the case with the only one non-zero charge, $T_0 \neq 0$.
The exact values of the $T_0$ are listed in Table~\ref{table:holethermo}
for the different cases.
\begin{table}[h]
  \centerline{
  \begin{tabular}{c|ccc}
    $(q_1,q_2,q_3)$  &   $(q,0,0)$  &  $(q,q,0)$  &  $(q,q,q)$  \\ \hline
    $T_0$            &   $1/\pi	$   &  $1/2\pi$   &  $0$		\\
    Asymp. Slope     &$\pi/\sqrt{2}$&  $\pi$      &  $2\pi$
  \end{tabular}
  }
  \caption{\footnotesize
    The list of the values of $T_0$ and the slops of the high
    temperature linear behavior for different cases.
  }
\label{table:holethermo}
\end{table}
Since the action is the difference between the black hole and
AdS solutions, the region of the phase diagram below $T_0$ is
completely undetermined.

In all the cases, the instability lines and the Hawking-Phage
phase transition lines merge at $T=T_0$ and $\mu = 1/R$ where
$R$ is the AdS curvature radius.
The solid instability lines rise linearly at temperature large
compared to $1/R$ and the slopes of the linear behavior is
also listed in Table~\ref{table:holethermo} for the different
cases.

We mention that the instabilities of the black holes
with $q_1=q_2$ and $q_1=q_2=q_3$ are obtained by assuming that
the equalities are subject to the thermal fluctuations
around the black hole saddle points.
Technically, this means that the Hessian of $I_\text{Gibbs}$
with respect to $q_j$ is computed by assuming they are the
independent parameters, and then the Hessian is evaluated
at the black hole saddle point with the relations
$q_1=q_2$ or $q_1=q_2=q_3$.
One may choose to consider the black hole with the
constraint $q_1 \equiv q_2$ or $q_1 \equiv q_2 \equiv q_3$
which is not subject to the fluctuations.
In this case, the action $I_\text{Gibbs}$ is obtained by
setting $q_1 \equiv q_2$ or $q_1 \equiv q_2 \equiv q_3$
first, and then the Hessian is computed with respect to
the single charge (and $r_H$).
The resulting thermodynamic system for this constrained case
is stable everywhere and the instability
lines are not observed.
The constraints reduce the dimension of the parameter space
and eliminate the direction of the instabilities.
This is the reason why the instability found in Reference~\cite{Cvetic:1999ne}
[with the configuration $(q,0,0)$] is not observed in the
analysis of Reference~\cite{Chamblin:1999tk}
[with the configuration $(q,q,q)$].

In concluding the review, we point out that the Gibbs free
energy $I_{\text{Gibbs}}$ is constructed out of the
quantities of the black hole.
The black hole is the solution of the equations of motion
and thus $I_{\text{Gibbs}}$ is made out of the on-shell quantities.
Therefore,
even though we expect the Gibbs free energy to be the
action of the theory in an infinitesimal neighborhood of
the black hole extremum, 
we have no guarantee that $I_{\text{Gibbs}}$ has any meaning
as the action of the theory away from the extremum.
The function $I_{\text{Gibbs}}$ is a local on-shell action in this
sense and it is not useful in examining the global
stability of the theory.
The examination of the global stability requires the
construction of the off-shell action that is valid
away from the extremum.

\section{Construction}\label{sec:construction}

\subsection{The Action,  the Metric, and the Topology}\label{sec:action}
We now embark on constructing
the reduced action of the system.
In this subsection, we first display the action
and the metric ansatz to start.
Then we assume the Euclidean black hole topology on
spacetime and deduce the resulting conditions on the metric.

\bigskip

As shown in Reference~\cite{Behrndt:1998jd},
the black hole solution reviewed in the previous section
can be obtained from the following special case of the
five dimensional $\mathcal{N}=2$ gauged
supergravity action
\begin{equation}\label{eq:action}
  I = - \frac{1}{8\pi G_5} \int d^5x\sqrt{g}
        \bigg[ \frac{1}{2}\mathcal{R} 
        - \frac{1}{2}\sum_{j=1}^3\frac{1}{2X_j^2}
                          (\partial_{\mu}X_j)(\partial^{\mu}X_j)
        - \frac{1}{4}\sum_{j=1}^3\frac{1}{2X_j^2}
                          F_{j\,\mu\nu}{F_j}^{\mu\nu}
        + \frac{V}{R^2} \bigg]
        \;,
\end{equation}
where only the bosonic sector is shown.
We have adopted the Euclidean signature and the sign of the action
is chosen so that the weight factor in the path integral appears as
$\exp (-I)$.
The Chern-Simons term is dropped from the action
because we will be considering
the electrostatic fields which have the form
$A_{j\,\nu} = A_{j\,\tau} \delta_{\nu,0}$, and the term identically
vanishes for such fields.
The determinant of the metric is denoted by $g$,
and $\mathcal{R}$ is the scalar curvature.
As in the previous section,
the fields $X_{j}(x)$ are the scalars and they satisfy
the constraint
\begin{equation}\label{eq:scalarConstraint}
  X_1(x)X_2(x)X_3(x) \equiv 1
        \;.
\end{equation}
The three gauged $U(1)^3\subset SO(6)$
Abelian fields $A_{j\,\nu}$ have the corresponding field strengths
\begin{equation}\label{eq:fieldStrength}
  F_{j\,\mu\nu} := A_{j\,\nu ; \mu} - A_{j\,\mu ; \nu}
      \;,
\end{equation}
where the indices $\mu$ and $\nu$ are the
spacetime indices and the semicolon denotes the
geometrical covariant derivative with respect to a
given metric, which, of course, can be replaced by
normal derivatives in this case.
The potential $V$ is defined as
\begin{equation}\label{eq:potential}
  V := 2 \sum_{j=1}^3 \frac{1}{X_j}
        \;.
\end{equation}
Finally in the action, the parameter $R$ is the length scale
of the theory which is related to the cosmological
constant of the asymptotic geometry.

\bigskip

We consider the static gravitational system with spherical symmetry
and thus take the metric ansatz
\begin{equation}\label{eq:metricAnsatz}
  ds^2 = b(r)^2 d\tau^2
	+ H(r)^{1/3} \big[ f(r)^{-1}dr^2 + r^2 d\Omega_3^2 \big]
	\;,
\end{equation}
where $\tau$ is the periodic Euclidean time whose period is set to
$2\pi R$.
Consequently, $b(r)$ is a dimensionless
function of $r$.
The form of the metric is completely general under the
static spherical symmetry but we choose
the particular form that resembles the solution of the previous
section in Equation (\ref{eq:metricSolution}).
This choice is motivated by the fact that we want the
resulting reduced action to reproduce the same form of the
solution as in the
previous section at the extremum.
Without loss of generality, we may assume that the dimensionless
function
$H(r)$ is positive definite and finite for finite values of
$r$ so that
the horizon in the Minkowski signature is determined
solely by the dimensionless function $f(r)$.
Notice that the static spherical symmetry requires
the scalar fields $X_j(x)$ be functions
only of $r$ as well.

In the computations that follow, we place the thermodynamic
system into a large box whose wall is located at $r=r_B$
(the word ``large'' is used with respect to the length
scale $R$).
In the AdS/CFT language, this is the ``cut-off boundary''
of the system and will be removed by taking the limit
$r_B\to\infty$.
The lower end of the range of the parameter $r$ 
is given at
$r=r_H$ which is the horizon radius of the black
hole in the Minkowski signature.
We will loosely use the terminology, ``horizon
radius'', even if we are discussing the Euclidean
black holes.

\bigskip

In addition to the static spherical symmetry,
we restrict our considerations to the action
that describes the spacetimes with Euclidean
black hole topology.
The Euclidean black hole topology of the spacetime
manifold is $\mathbb{R}^2\times S^3$
and the boundary of the manifold has the topology $S^1\times S^3$
where the $S^1$ factor corresponds to the periodic Euclidean time
circle.
In the actual computation, we assume the topology $S^1\times S^3$
at the wall of the box, $r=r_B$, with the limit $r_B\to\infty$
in our mind.

We first want to impose the topology
$\mathbb{R}^2\times S^3$ on the entire spacetime.
However, the Euler characteristic of any odd dimensional
manifold identically vanishes and it is not desired here.
Thus, we fix one of the angular coordinates in $S^3$ and
require this slice to have the topology $\mathbb{R}^2\times S^2$
with the Euler characteristic $\chi$ equaling $2$.
This yields the condition%
\footnote{
  The topological condition imposed here is the same as the one
  in Reference~\cite{Braden:1990hw} and we have borrowed the
  result of the reference with a straightforward modification
  for the form of the metric that we have adopted.
}
\begin{equation}\label{eq:topologyHorizon}
  \Big[
        f H^{-1/3} \big\{ (r H^{-1/6} )' \big\}^2
  \Big]_{r=r_H}
  = 0
        \;,
\end{equation}
where the prime denotes the derivative with respect to the
radial coordinate $r$
and the left-hand side of the equation is evaluated at
$r=r_H$.
We will see that this condition boils down to the expected
condition for a black hole horizon, $f(r_H)=0$.

Next, consider the
$\mathbb{R}^2$ part of the spacetime.
This  is commonly called
``the cigar'' where the tip of the cigar corresponds to the
black hole horizon and thus it is
parametrized by the lower end of the radial coordinate
$r=r_H$, whereas the other end of the cigar is the boundary
$S^1$.
In this topology, the $S^1$ time circle degenerates
at the tip $r=r_H$ and thus we have
\begin{equation}\label{eq:b_rH_zero}
  b(r_H) = 0
        \;.
\end{equation}
Also this topology
requires the $\tau$-$r$ slice of the spacetime to have
the Euler characteristic $\chi = 1$ as the cigar
is topologically a two dimensional ball with boundary.
This condition parallels with the usually known ``smoothness
condition'' at the tip of the cigar.%
\footnote{
  Most commonly, the smoothness is achieved by setting the
  period of the $S^1$ time circle to a specific value.
  Though equivalent, we are fixing the period to $2\pi R$
  and require the metric to satisfy a certain condition.
}
Being cautious for the orientation of the $S^1$
time circle at the wall of the box,
the Gauss-Bonnet theorem yields
\begin{equation}\label{eq:cigarCondition}
  \left[ b' f^{1/2} H^{-1/6} \right]_{r=r_H} = 1/R
  \;.
\end{equation}
There are terms evaluated at $r=r_B$, one from the Gauss curvature part
and the other from the geodesic curvature of the boundary curve, 
but they cancel and absent in the above expression.

\bigskip

\subsection{The Boundary Data}\label{sec:boundaryData}
We encode the boundary data of the system in this subsection.
Those data will be the fixed parameters of the ``histories''
of the paths in
the path integral and they will not be varied in the
extremization of the action.
Since we want to consider and fit
the black hole system in the context of the AdS/CFT correspondence, 
we have the metric satisfy
the asymptotic AdS geometry near the boundary.
Also we are interested in the grand canonical
ensemble of the black hole thermodynamics, so the
temperature and chemical potentials are the fixed parameters.
If there is a physically equivalent field theory on
the boundary, then those fixed black hole thermodynamic
data should be identical to the ones in the field theory.
Therefore, it is natural to encode the fixed parameters
on the boundary where the geometry is fixed and the
corresponding field theory is supposed to live.

\bigskip

First, we want the asymptotic AdS geometry near the boundary.
We impose the asymptotic conditions on the
metric ansatz (\ref{eq:metricAnsatz})
\begin{equation}\label{eq:asymptAdSCond}
  b(r_B) \to \ell \frac{r_B}{R}
        \;,
        \quad
  f(r_B) \to \frac{r_B^2}{R^2}
        \;,
        \quad
  H(r_B) \to 1
        \;,
        \quad
        \text{as}
        \quad
  r_B \to \infty
        \;.
\end{equation}
The parameter $\ell$ is an arbitrary dimensionless positive constant.
The motivation for introducing the parameter $\ell$ shall
be given shortly in the discussion of the temperature below.
One sees that with the analytic continuation,
$t = -i \ell\tau$, the metric ansatz (\ref{eq:metricAnsatz})
approaches to the AdS metric with the curvature radius
$R$, up to the first order in $r_B$.

\bigskip

Next we discuss the temperature of the system.
It is sometimes seen that an observer is not specified
for the discussion of the black hole temperature.
However, the inverse temperature is the circumference
of the $S^1$ Euclidean time circle and it depends 
on the location of the observer.
For our metric, the local, physical inverse
temperature at $r=r_B$
is the proper length of the $S^1$ circumference
\begin{equation}\label{eq:betaTilde}
	\tilde\beta = \int^{2\pi R}_{0} d\tau b(r_B) = 2 \pi R \, b(r_B)
	\;.
\end{equation}
Due to the asymptotic AdS condition on $b(r)$ in
(\ref{eq:asymptAdSCond}),
the physical temperature is red-shifted to zero as
the wall is pushed to the boundary,
{\it i.e.}, as $r_B \to \infty$.
This fact has been known since as early as Hawking
and Page \cite{Hawking:1982dh}.
We could have discussed the thermodynamics defined in
a finite box with $r_B<\infty$, but since we are
interested in the AdS/CFT correspondence,
we must take the limit $r_B\to\infty$.

In discussing the thermodynamics of AdS black holes
in the infinite box,
the unphysical temperature, so-called ``renormalized temperature'',
has been used.
(Among the works that adopt the ``renormalized'' temperature,
we refer to one paper \cite{Brill:1997mf} in which this issue
is clearly addressed.)
The ``renormalized'' temperature is defined
by the product of the physical temperature at $r=r_B$ and
the red-shift factor of the black hole geometry
so that the limit $r_B\to\infty$ yields a finite value.
(Consequently, if we set the boundary temperature to the finite
value, the temperature diverges everywhere in the bulk).
Even though the physical meaning of the redefined temperature
has remained unclear,
there have been indications that this, at least formally,
is the correct temperature
to be used on the boundary.
The indications include the correct 
forms of the heat capacity,
first law of thermodynamics 
and the quarter area law of the entropy,
all with the redefined temperature \cite{Hawking:1982dh,Brill:1997mf,Brown:1994gs}.
In addition, our analysis below also requires a
similar kind of redefinition in the temperature.
Therefore, we introduce the new temperature $\beta$
defined at the boundary by
\begin{equation}\label{eq:boundaryTemp}
	\beta := \lim_{r_B\to\infty}
	          \bigg( \frac{R}{r_B} \bigg) \tilde\beta
					= 2 \pi \ell R
        \;.
\end{equation}
Here we see that the temperature is directly proportional
to the arbitrary parameter $\ell$.
Thus the temperature is a thermodynamic quantity that
{\it we} can tune, not the system, as a grand canonical
ensemble should be.
We will discuss more on the redefined temperature in
Section~\ref{subsec:adsCftBoundaryTemp}.

\bigskip

Finally, we encode the chemical potentials that are
conjugate to the $R$-charges of the system.
In a grand partition function, a chemical potential
couples to the conjugate charge which is the time
component of a conserved current.
Therefore, the chemical potential is directly related
to the time component
of the gauge field that minimally couples to the conserved current.
Thus we define the chemical potentials at the wall
$r=r_B$ as
\begin{equation}\label{eq:defMuTilde}
  \tilde\mu_j := \frac{1}{\tilde\beta}\ln 
                \exp \bigg[ i \oint A_j / R \bigg]
        \;,
\end{equation}
where the integral on the right-hand side is defined on
the $S^1$ time circle at $r=r_B$ and $A_j$ are the Hermitian
gauge field one-forms.
In the definition, the length scale $R$ is inserted for
the dimensional reason.
We choose the (Wick rotated) gauge field of the form
\begin{equation}\label{eq:gaugeFields}
	A_{j\,\nu} = A_{j\,\tau}(r) \delta_{\nu,0}
	\;.
\end{equation}
This form assumes the electrostatic spherical symmetry.
Then we have
\begin{equation}\label{eq:continuationA}
  \oint A_j = \oint (-i \ell A_{j\,t}) (i dt/\ell) 
            = \oint A_{j\,\tau} d\tau
            = 2 \pi R A_{j\,\tau} (r_B)
        \;,
\end{equation}
where the analytic continuation $t=-i\ell\tau$ has been
done and $A_{j\,\tau}$ now is pure imaginary.
Therefore, we get
\begin{equation}\label{eq:muTilde}
  \tilde\mu_j  = \frac{2\pi i}{\tilde\beta} A_{j\,\tau}(r_B)
        \;.
\end{equation}

Now if we assume that the gauge fields $A_{j\,\nu}(r_B)$
are finite as $r_B\to\infty$,%
\footnote{
  Alternatively, we may assume $A_{j\,\mu}(r_B)$ are singular
  so that $\tilde\mu_j$ remain finite.
  This can also lead to the same conclusion at the end.
}
then the chemical potentials $\tilde\mu_j$ at the boundary
vanish because $\tilde\beta$ diverges
due to the asymptotic AdS condition
(\ref{eq:asymptAdSCond}).
Thus, similar to the case of the temperature, we define
the boundary chemical potentials by
\begin{equation}\label{eq:boundaryChemp}
  \mu_j := \lim_{r_B\to\infty} 
                \bigg( \frac{r_B}{R} \bigg) \tilde\mu_j
        \;.
\end{equation}
This definition will be discussed further in
Section~\ref{subsec:adsCftBoundaryTemp} along with the
temperature $\beta$ defined in Equation~(\ref{eq:boundaryTemp}).

We add that since we have the condition (\ref{eq:b_rH_zero}),
we need to have the conditions
\begin{equation}\label{eq:Azero}
	A_{j\,\tau} (r_H) = 0
	\;,
\end{equation}
so that the proper local value of the gauge potential
$A_{j\,\tau}(r_H)/b(r_H)$ can be set to zero as the reference
point of the potential.

\subsection{The Constraints}\label{sec:constraints}
Our system has two sets of constraints;
Gauss' law and the Hamiltonian constraint.
The physical states of the theory must satisfy the constraints
and we treat them fixed and not subject to the variation.
The way they arise is very similar.%
\footnote{
  See Wald's textbook \cite{book:Wald} for a pedagogical exposition
  of the constraints.
  It is also instructive to observe how the Gauss law and the
  Hamiltonian constraint are identically treated in 
  References~\cite{Gross:1980br} and \cite{Gross:1982cv},
  respectively.

  One might be slightly puzzled by the relation between the dynamical Einstein equations and the Hamiltonian constraint
  in a static spacetime manifold,
  because the constraint is the initial value condition for the
  Cauchy development problem.
  The Hamiltonian constraint in a static spacetime determines the
  geometry of space-like leaves of the spacetime foliation but it
  does not determine the way those leaves should be ``stack up''
  in the time-like direction.
  In other words, the Hamiltonian constraint cannot determine the
  function $b(r)$ in the metric (\ref{eq:metricAnsatz}) and the
  determination of the function requires the full Einstein equations.
}
Among the Maxwell and Einstein equations, the constraints correspond
to the non-dynamical equations, {\it i.e.}, the ones without second
order time derivatives.
The constraints can also be viewed as the resulting equations of
the variation with respect to Lagrange multipliers.
In the electrodynamics, the Lagrange multiplier is the time component
of the gauge field.
In the ADM formalism \cite{ADM,book:Wald}
of the geometrodynamics, the shift and lapse functions are the
Lagrange multipliers.
Below, we bring the action (\ref{eq:action}) to the form where the
Lagrange multipliers explicitly appear and eliminate the constraints
from the action, without imposing the dynamical equations of motion.

\bigskip

We first consider the gravitational part of the action
(\ref{eq:action}),
\begin{equation}
  I_G = - \frac{1}{8\pi G_5} \int d^5x \sqrt{g}\,
        \frac{1}{2}\mathcal{R}
        \;.
\end{equation}
One can verify that
with the metric of the form (\ref{eq:metricAnsatz}),
the scalar curvature $\mathcal{R}$ can be written as
\begin{equation}
  - \frac{1}{2} \mathcal{R} = 
        \frac{ f^{1/2} }{ b \, r^{3} H^{2/3} }
        \big( b' r^3 f^{1/2} H^{1/3} \big)'
        + {G^\tau}_\tau
        \;,
\end{equation}
where the last term is the ${(^\tau}_\tau)$-component of the
Einstein tensor and it explicitly has the form
\begin{equation}\label{eq:Gtautau}
  {G^\tau}_\tau = \frac{3}{2 r^3 H^{1/2} (r H^{1/6})'}
           \Big[ r^2 H^{1/3} \Big(fH^{-1/3}\big\{(rH^{1/6})'\big\}^2-1\Big) \Big]'
        \;.
\end{equation}
Notice that this expression is independent of the function $b$.
Now $I_G$ can be expressed as
\begin{equation}\label{eq:IGravity}
  I_G = \frac{\omega_3}{8\pi G_5} \, 2\pi R
        \bigg\{
           \int dr \, ( b' r^3 f^{1/2} H^{1/3} )'
          + \int dr \, b \, r^3 f^{-1/2} H^{2/3} {G^\tau}_\tau
        \bigg\}
        \;.
\end{equation}

To proceed further, we need to introduce the
Gibbons-Hawking term \cite{Gibbons:1976ue} to the
gravitational part of the action $I_G$.
In general, the Gibbons-Hawking term is required
so that a gravitational action has the consistent
variational principle
with respect to general variation of the metric
whose induced metric on the wall of the box
is held fixed.%
\footnote{
  If one considers more specific kind of the metric
  variation, namely, the metric whose induced
  metric on the wall of the box is held fixed {\it and}
  the first derivatives of the metric on the wall
  are also fixed, then the Gibbons-Hawking surface
  term is absent.
}
The Gibbons-Hawking term is not unique and one is allowed
to include arbitrary term that depends only on the induced
metric on the wall which can be immersed in a different
spacetime.
This arbitrary term is commonly used to set the reference
background geometry of the action.
However, the introduction of the background geometry
is arbitrary and sometimes it is not possible to set
a seemingly natural background.
Thus, it is more desirable to construct our
action independent of the background,
therefore, we do not include the ambiguous term here
and instead, adopt the AdS/CFT motivated background
independent counterterms
as discussed in the next subsection.
The Gibbons-Hawking term without the subtraction term
has the form
\begin{equation}\label{eq:GibbonsHawking}
	I_{GH} = \frac{1}{ 8 \pi G_5 } \int_{\partial M} d^4x
		\sqrt{\gamma} K
	\;,
\end{equation}
where $\partial M$ is the wall at $r=r_B$ with the metric
\begin{equation}\label{eq:4dmetric}
	ds_4^{2} = b(r_B)^2 d\tau^2 + H(r_B)^{1/3} r_B^2 d\Omega_3^2
	\;,
\end{equation}
and $\gamma$ is the determinant with respect to this metric.
The factor $K$ is the trace of the extrinsic curvature
\begin{equation}
	K_{\mu\nu} = - \frac{1}{2} \left( n_{\mu;\nu} + n_{\nu;\mu} \right)
	\;,
\end{equation}
where $n_\mu$ is a normal unit vector of the wall, $\partial M$,
and is given by
\begin{equation}
	n_\mu = \Big( 0, H^{1/6}/f^{1/2}, 0, 0, 0 \Big)
	\;.
\end{equation}
A straightforward computation yields
\begin{equation}\label{eq:newGH}
	I_{GH} = - \frac{ \omega_3 }{ 8 \pi G_5 } 2 \pi R
		\Big[
		f^{1/2} H^{-1/6} \left( b \, r^3 H^{1/2} \right)'
		\Big]_{r = r_B}
	\;.
\end{equation}

We combine Equations (\ref{eq:IGravity}) and (\ref{eq:newGH}).
After some algebra, we obtain
\begin{align}
  I_G + I_{GH} =& - \frac{ \omega_3 }{ 8\pi G_5 }
        \bigg\{ \tilde\beta \,
                \Big[ f^{1/2} H^{-1/6} ( r^3H^{1/2} )' \Big]_{r=r_B}
                + 2 \pi r_H^3 \sqrt{ H(r_H) }
        \bigg\}
        \nonumber\\
        &+ \frac{ \omega_3 }{ 8\pi G_5 } \, 2\pi R
                \int dr \, b \, r^3 f^{-1/2} H^{2/3} {G^\tau}_\tau
        \;,
\end{align}
where the topological condition (\ref{eq:cigarCondition}) 
and the definition of $\tilde\beta$ in (\ref{eq:betaTilde})
have been used.

We proceed to the Maxwell part of the action
\begin{equation}
  I_{EM} = \frac{1}{8\pi G_5} \int d^5 x \sqrt{g} \,
           \frac{1}{4} \sum_j \frac{1}{2X_j^2} \,
           F_{j\, \mu\nu}{F_j}^{\mu\nu}
           \;.
\end{equation}
We define the Maxwell field tensors as in
Equation~(\ref{eq:fieldStrength})
and with our metric (\ref{eq:metricAnsatz}) and the 
gauge fields (\ref{eq:gaugeFields}),
the only nonzero components are
\begin{equation}\label{eq:Fs}
  {F_{j\, 0 1}} = - {F_{j\, 1 0}} = - ({A_{j\, \tau}})'
        \;,
\end{equation}
where the components with $\mu, \nu = 1$ are
the radial $r$-component.
We introduce auxiliary fields $E_j(r)$ such that
\begin{equation}
  I_{EM} = \frac{\omega_3}{8\pi G_5} \pi R
           \int dr \, b^{-1} r^3 f^{1/2} H^{1/3}
           \sum_j \frac{1}{2X_j^2} 
           ( - E_j^2 + 2 E_j F_{j\, 10})
           \;.
\end{equation}
Notice that $\delta I_{EM}/\delta E_j = 0$ yields
$E_j = F_{j\, 10}$ and insertion of this equation
back into $I_{EM}$ recovers the original form of the
Maxwell action.
For the factor $F_{j\, 10}$ in the second term,
we use the relation (\ref{eq:Fs}) and express that
in terms of the derivative with respect to $r$ acting
on $A_{j\,\tau}$.
We then integrate this by parts to obtain
\begin{align}\label{eq:IEMSplit}
  I_{EM} =& - \frac{\omega_3}{8\pi G_5} \pi R
          \int dr \, b^{-1} r^3 f^{1/2} H^{1/3}
          \sum_j \frac{1}{2X_j^2} E_j^2
          \nonumber\\
          &- \frac{\omega_3}{8\pi G_5} 2\pi R
          \int dr \sum_j \Big( b^{-1} r^3 f^{1/2} H^{1/3}
                    \frac{1}{2X_j^2} E_j \Big)' A_{j\,\tau}
          \nonumber\\
          &+ \frac{\omega_3}{8\pi G_5} 2 \pi R
             \int dr \Big( b^{-1} r^3 f^{1/2} H^{1/3}
                   \sum_j \frac{1}{2X_j^2} E_j \, A_{j\,\tau} \Big)'
\end{align}
We note that the last term is a total derivative and does not
contribute to the equations of motion.

Finally, the scalar part of the action takes the form
\begin{align}
  I_S =& - \frac{1}{8\pi G_5} \int d^5 x \sqrt{g} \,
        \sum_j \bigg[
          - \frac{1}{4X_j^2}
            (\partial_\mu X_j) (\partial^\mu X_j)
            + \frac{2}{R^2}\frac{1}{X_j}
        \bigg]
        \nonumber\\
        =&- \frac{\omega_3}{8\pi G_5} \, 2\pi R
        \int dr \, b \, r^3 f^{-1/2} H^{2/3}
          \sum_j \bigg[
            - f H^{-1/3} \, \frac{(X_j')^2}{4X_j^2}
            + \frac{2}{R^2} \frac{1}{X_j}
          \bigg]
          \;.
\end{align}

Now in the total action
\begin{equation}
  I = I_G + I_{GH} + I_{EM} + I_S
  \;,
\end{equation}
we clearly see that $A_{j\,\tau}$ in $I_{EM}$ 
are the Lagrange multipliers.
The variation with respect to those fields yields
Gauss' law constraint of the system,
\begin{equation}\label{eq:gaussLaw}
  \Big(
    b^{-1} r^3 f^{1/2} H^{1/3} \frac{1}{2X_j^2} E_j
  \Big)' = 0
      \;,
\end{equation}
where the repeated index $j$ is {\it not} summed over.
We introduce the integration constants $-ie_j$ so that
\begin{equation}\label{eq:electricField}
  E_j = -2i \, b \, f^{-1/2} H^{-1/3} X_j^2 \;
        \frac{e_j}{r^3}
        \;.
\end{equation}
The constants $e_j$ describe the real (as opposed to
complex) charges that satisfy Gauss' law and the
factor of $-i$ is inserted to account for the analytic
continuation from Minkowski to Euclidean space.

Using Equation~(\ref{eq:electricField}), $I_{EM}$ can
be expressed as
\begin{equation}
  I_{EM} = \frac{\omega_3}{8\pi G_5} \bigg\{ 2\pi R
        \int dr \, b \, r^{-3} f^{-1/2} H^{-1/3}
        \sum_j e_j^2 X_j^2
        - 
        R^3\tilde\beta \sum_j \Big( \frac{e_j}{R^2} \Big) \tilde\mu_j
        \bigg\}
        \;,
\end{equation}
where the last term comes from the evaluation of the
total derivative term in Equation~(\ref{eq:IEMSplit}) and
we have applied the boundary conditions
(\ref{eq:muTilde}) and (\ref{eq:Azero}).

The total action at this point has the form
\begin{align}\label{eq:preTotalAction}
  I =& - \frac{ \omega_3 R^3 }{ 8\pi G_5 }
        \bigg\{ \frac{\tilde\beta}{R^3} \,
                \Big[ f^{1/2} H^{-1/6} ( r^3H^{1/2} )' \Big]_{r=r_B}
                + \tilde\beta \sum_j \Big( \frac{e_j}{R^2} \Big) \tilde\mu_j
                + 2 \pi \Big( \frac{r_H}{R} \Big)^3 \sqrt{ H(r_H) }
        \bigg\}
        \nonumber\\
        &+ \frac{ \omega_3 }{ 8\pi G_5 } \, 2\pi
            \int dr \, b \, r^3 f^{-1/2} H^{2/3}
              \Big( {G^\tau}_\tau - 8\pi G_5 {T^\tau}_\tau \Big)
        \;,
\end{align}
where ${T^\tau}_\tau$ is defined as
\begin{equation}\label{eq:energyMomentumTensor}
  {T^\tau}_\tau := \frac{1}{8\pi G_5}
	\sum_j \bigg[
                - \frac{ e_j^2 \, X_j^2 }{ r^6 H }
                - \frac{1}{4} \frac{ f \, (X_j')^2 }{H^{1/3}X_j^2}
                + \frac{2}{R^2}\;\frac{1}{X_j}
        \bigg]
        \;.
\end{equation}
We have noted that ${G^\tau}_\tau$ shown in (\ref{eq:Gtautau}) is free
of the function $b$.
Therefore, in the total action (\ref{eq:preTotalAction}), we
explicitly see that the function $b$ is the Lagrange multiplier
that enforces the constraint
\begin{equation}\label{eq:hamilConstraint}
  {G^\tau}_\tau = 8\pi G_5 {T^\tau}_\tau
        \;.
\end{equation}
One can verify that this, in fact, is the ${(^\tau}_\tau)$-component
of the Einstein equations and thus, 
this is the Hamiltonian constraint of the system.%
\footnote{
  In general, the Hamiltonian constraint of a gravitational
  system is given as
  $
	G_{\mu\nu}n^\nu = 8\pi G_5 T_{\mu\nu}n^{\nu}
  $,
  where $n^\nu$ is a unit time-like vector.
  In our metric (\ref{eq:metricAnsatz}) and action (\ref{eq:action}), there
  is no non-vanishing off-diagonal components in the Einstein and
  energy-momentum tensors.
  Therefore, the Hamiltonian constraint of our system is
  $
	{G^\tau}_\tau = 8\pi G_5 {T^\tau}_\tau
  $
  which is the ${(^\tau}_\tau)$-component of the Einstein equation.

  The variation of the action (\ref{eq:action}) with respect
  to the metric yields the Einstein equations of the system
  \begin{equation}
    G_{\mu\nu} = \;
      \sum_j\frac{1}{2X_j^2}
        \bigg\{ (\partial_\mu X_j)(\partial_\nu X_j)
           - \frac{1}{2}g_{\mu\nu}(\partial_\lambda X_j)(\partial^\lambda X_j)
        - \frac{1}{4}g_{\mu\nu}F_{j\, \alpha\beta}{F_j}^{\alpha\beta}
         + {F_{j\, \mu}}^\lambda F_{j\,\nu\lambda}
           \bigg\}
         + g_{\mu\nu} \frac{V}{R^2}
        \;.
  \end{equation}
  From this equation, it is easy to verify that
  Equation~(\ref{eq:hamilConstraint}) with (\ref{eq:energyMomentumTensor})
  is the ${(^\tau}_\tau)$-component of the Einstein equations.

  The Einstein equations of the theory
  is frequently presented in the form that does not have the
  scalar curvature.
  Expressing the Einstein equations in this form
  requires to take the trace of the Einstein equation
  and this assumes
  the equality of other components than the constraint equation.
  Therefore, it is not adequate for extracting only the constraint
  equation.
}

\bigskip

Given the expressions (\ref{eq:Gtautau}) and (\ref{eq:energyMomentumTensor}),
the Hamiltonian constraint (\ref{eq:hamilConstraint}) is a
differential equation of functions $f$, $H$ and $X_j$ with respect
to the variable $r$.
We solve this differential equation by making ansatz which is
similar to the one made in Reference~\cite{Behrndt:1998jd},
\begin{equation}\label{eq:fXansatz}
  f(r) = 1 - \frac{r_0^2}{r^2} + \frac{r^2}{R^2} H(r)
  \quad\text{and}\quad
  X_j(r) = H_j^{-1} H(r)^{1/3}
  \;,
\end{equation}
where $r_0$ is an undetermined parameter and
the functions $H_j$ are defined as $H_j:=1+q_j/r^2$.
Now because of the constraint (\ref{eq:scalarConstraint}) imposed
on the scalar fields, the function $H(r)$ is determined to be
\begin{equation}\label{eq:HAnsatz}
  H(r) = H_1(r)H_2(r)H_3(r)
        \;.
\end{equation}
Thus we see that $q_j$ are the three parameters of the three 
fields; $H$ and $X_j$ with the constraint
(\ref{eq:scalarConstraint}).
Notice that this ansatz is consistent with the asymptotic
condition (\ref{eq:asymptAdSCond}).
The form of the function $H$ above implies that the topology
condition (\ref{eq:topologyHorizon}) is equivalent to
the equation
\begin{equation}\label{eq:horizonCondition}
  f(r_H) = 0
        \;,
\end{equation}
as advertised.
Using this condition, we can trade the parameter $r_0$ of the function
$f$ with the parameter $r_H$,
just as in Equation~(\ref{eq:m}).
Then the Hamiltonian constraint (\ref{eq:hamilConstraint})
is satisfied if and only if
\begin{equation}\label{eq:newEQRel}
  e_j^2 = q_j ( r_H^2 + q_j)
        \bigg[
                1 + \frac{r_H^2}{R^2} \prod_{k\neq j}H_k(r_H)
        \bigg]
        \;,
\end{equation}
where the repeated index $j$ is not summed over.
One observes that these relations are identical
to the ones obtained in the on-shell analysis
of Section~\ref{sec:review}
as shown in Equation~(\ref{eq:eqRelations}).

\bigskip

In solving the Hamiltonian constraint, we have adopted the rather
tight ansatz (\ref{eq:fXansatz}).
This narrow assumption on the functions restrict the form of
possible solutions.
Instead of adopting the ansatz discussed above, we can assume
the forms of the scalars $X_j$ and the function $H$ only, and
regard the Hamiltonian constraint as a first order differential
equation of the function $f(r)$.
The solution to the differential equation generates a new
family of solutions that includes
the solution discussed in Section~\ref{sec:review} as a special case.
We will discuss this more general solution in
Section~\ref{subsec:moreGeneralSolution}.

\subsection{The Boundary Reduced Action}\label{sec:boundaryAction}
In the previous subsection, we have solved and eliminated the
constraints of the system.
Thus the action now takes the form
\begin{equation}
  I = - \frac{ \omega_3 R^3 }{ 8\pi G_5 }
        \bigg\{ \frac{\tilde\beta}{R^3} \,
                \Big[ f^{1/2} H^{-1/6} ( r^3H^{1/2} )' \Big]_{r=r_B}
                + \tilde\beta \sum_j \Big( \frac{e_j}{R^2} \Big) \tilde\mu_j
                + 2 \pi \Big( \frac{r_H}{R} \Big)^3 \sqrt{ H(r_H) }
        \bigg\}
       \;.
\end{equation}
This action describes the thermodynamics of the black hole system
in the box with the wall at $r=r_B$.
As a grand canonical ensemble, the temperature
$\tilde\beta$ and the chemical potentials $\tilde\mu_j$ are
defined at the wall and
the grand canonical data at other locations in
the box cannot be deduced from $\tilde\beta$ and
$\tilde\mu_j$, for the functions $b$ and $A_{j\,\tau}$
are the Lagrange multipliers and
unspecified except on shell.%
\footnote{
  This can be physically understood.
  Imagine an experiment with a thermodynamic system in a box
  where the number of electrically charged particles 
  of the system is allowed to vary.
  An experimenter sets up a temperature on the wall of the box
  using a kind of heat reservoir and also applies an electric
  voltage on the wall.
  The experimenter cannot define the temperature and the chemical
  potential of the system as a whole
  until the system reaches the thermodynamic equilibrium.
  The analogy is clear.
}

We are interested in the thermodynamics with the limit $r_B\to\infty$.
The action has the asymptotic form
\begin{align}\label{eq:divergentAction}
  I =& \frac{\omega_3}{8\pi G_5} \bigg( \tilde\beta \frac{R}{r_B} \bigg)
        \bigg[ E R^3
          -3 \frac{r_B^4}{R^2} 
          - \frac{3}{2} \Big( 1 + \sum_j \frac{q_j}{R^2} \Big) r_B^2
          - \frac{1}{4} \sum_j q_j
          \nonumber\\
          &+\frac{1}{24R^2}\Big\{\sum_j q_j^2 
                             -10( q_1 q_2 + q_2 q_3 + q_3 q_1 )\Big\}
          + \mathcal{O}\Big( \frac{1}{r_B^2} \Big)
        \bigg]
        \nonumber\\
        &-\frac{\omega_3 R^3}{8\pi G_5}
         \bigg\{\tilde\beta \sum_j \Big( \frac{e_j}{R^2} \Big) \tilde\mu_j
                + 2 \pi \Big( \frac{r_H}{R} \Big)^3 \sqrt{ H(r_H) }
        \bigg\}
        \;,
\end{align}
where we have defined
\begin{equation}\label{eq:defE}
  E:= \frac{1}{R^3} \bigg( \frac{3}{8} R^2 + \frac{3}{2} r_0^2 + \sum_j q_j \bigg)
  \;,
\end{equation}
with $r_0$ as appearing in Equation~(\ref{eq:m}).

The second and third terms in the action are divergent
as the wall approaches the boundary.
Recall that we did not include the arbitrary background subtraction
term in the Gibbons-Hawking term (\ref{eq:GibbonsHawking}).
This is because we prefer to regulate the divergence of the action by
using the AdS/CFT-motivated counterterm technique
\cite{Henningson:1998gx,Balasubramanian:1999re,Emparan:1999pm}
without arbitrarily assuming a background geometry.
Therefore, we employ the standard counterterms
\begin{equation}
	I_1 = \frac{1}{ 8 \pi G_5 } \int_{ \partial M } d^4 x
		\sqrt{\gamma} \frac{3}{R}
	\;,\quad
	I_2 = \frac{1}{ 8 \pi G_5 } \int_{ \partial M } d^4 x
		\sqrt{\gamma} \frac{R}{4}\mathcal{R}_4
	\;,
\end{equation}
where $\mathcal{R}_4$ is the scalar curvature with respect to the four-metric
(\ref{eq:4dmetric}) and we have
\begin{equation}
	\mathcal{R}_4 = \frac{6}{r_B^2 H(r_B)^{1/3}}
	\;.
\end{equation}
%
%
The evaluation of the counterterms is straightforward and yields
the asymptotic form
\begin{align}
	I_1 + I_2 =& \frac{ \omega_3 }{ 8 \pi G_5 }
	             \bigg( \tilde\beta \frac{R}{r_B} \bigg)
          \bigg[
                3\frac{r_B^4}{R^2}
                + \frac{3}{2}
                   \Big( 1 + \sum_j \frac{ q_j }{R^2} \Big) r_B^2
                + \frac{1}{4} \sum_j q_j
        \nonumber\\
                &- \frac{3}{8} \frac{1}{R^2}
                           \Big\{
                             \sum_j q_j^2 - 2 (q_1q_2+q_2q_3+q_3q_1)
                           \Big\}
		   + \mathcal{O}\Big( \frac{1}{r_B^2} \Big)
          \bigg]
	\;.
\end{align}
We see that the first two terms
in the above expression cancel the divergent
terms in Equation~(\ref{eq:divergentAction}).

In addition to the standard counterterms just computed, we
will find that we need to include the ``finite counterterm''
discussed by Liu and Sabra \cite{Liu:2004it}, so that the
resulting reduced action correctly reproduces the black
hole solution of Section~\ref{sec:review}.
Liu and Sabra argued that in the
presence of matter fields, it is possible to consider
other forms of the counterterms than the standard ones.
The possible new counterterms include
\begin{equation}\label{eq:finiteCounterterm}
  I_\text{finite} = \frac{1}{8 \pi G_5} \int_{\partial M} d^4x
                   \sqrt{\gamma} \, \vec\phi \cdot \vec\phi
        \;,
\end{equation}
where the scalar fields $\vec\phi = (\phi_1, \phi_2)$
are related to $X_j$ \cite{Cvetic:1999xp}, {\it via}
\begin{equation}
  X_j = \exp \Big[-\frac{1}{2} \, \vec a_j \cdot \vec\phi \Big]
        \;,
\end{equation}
and the vectors $\vec a_j$ can be chosen as
\begin{equation}
  \vec a_1 = \Big( \frac{2}{\sqrt{6}} , \sqrt{2} \Big)
        \;,\;
  \vec a_2 = \Big( \frac{2}{\sqrt{6}} , - \sqrt{2} \Big)
        \;,\;
  \vec a_3 = \Big( - \frac{4}{\sqrt{6}} , 0 \Big)
        \;.
\end{equation}
Under the assumption of the spherical symmetry, this is
the only non-vanishing new counterterm.
Though this term is finite as the limit, $r_B\to\infty$,
is taken, the inclusion of this term produces the mass
consistent with the one reported in Reference~\cite{Behrndt:1998jd}.
To avoid diverting the topic here, we yield the further
discussion on the finite counterterm to
Section~\ref{subsec:finiteCounterterm}.
With the forms shown in Equations (\ref{eq:fXansatz}) and (\ref{eq:HAnsatz}), the finite counterterm 
(\ref{eq:finiteCounterterm}) is
\begin{equation}
  I_\text{finite} = \frac{ \omega_3 }{ 8 \pi G_5 }
                \bigg( \tilde\beta \frac{R}{r_B} \bigg)
                \frac{1}{3R^2}
                \Big\{ \sum_j q_j^2
                        -  (q_1q_2 + q_2q_3 + q_3q_1) \Big\}
                + \mathcal{O}\Big( \frac{1}{r_B^3} \Big)
        \;.
\end{equation}

Combining all the results together, we have the action
\begin{equation}
  I = \frac{\omega_3 R^3}{8\pi G_5} \bigg[
             \bigg( \tilde\beta \frac{R}{r_B} \bigg)
             \Big\{ E 
              - \sum_j \Big( \frac{e_j}{R^2} \Big) 
                        \Big( \frac{r_B}{R} \Big) \tilde\mu_j \Big\}
              - 2 \pi \Big( \frac{r_H}{R} \Big)^3 \sqrt{ H(r_H) }
              + \mathcal{O} \Big( \frac{1}{r_B^3} \Big)
              \bigg]
              \;.
\end{equation}
Recall from Section~\ref{sec:boundaryData} that the inverse temperature
$\tilde\beta$ diverges as the limit $r_B\to\infty$ is taken.
The form of the action urges us to redefine the temperature as
in Equation~(\ref{eq:boundaryTemp}) and similarly for the
chemical potentials as in Equation~(\ref{eq:boundaryChemp}).
Adopting the definitions, the limit $r_B\to\infty$ yields
the action
\begin{equation}\label{eq:reducedAction}
  I^* = \beta \bigg\{ E - \sum_j \Big( \frac{e_j}{R^2} \Big) \mu_j \bigg\}
        - 2\pi \Big( \frac{r_H}{R} \Big)^3 \sqrt{H(r_H)}
        \;,
\end{equation}
where in the definition, the overall dimensionless
constant $\omega_3R^3/8\pi G_5$
has been absorbed [commonly in the literature, the convention
$G_5=(\pi/4)R^3$ is adopted for the same effect]
and the parameters $q_j$ and $e_j$ are assumed to be
related through Equation~(\ref{eq:newEQRel}).
We recall that the temperature $\beta$ and $\mu_j$ are 
{\it not} the physical quantities in the usual sense,
and $I^*$ is an action that is defined
only on the boundary.
In this sense, this is the boundary reduced action derived from
the bulk black hole thermodynamic system.

We end this section by commenting on a few expected properties of
the action.
First, we note that extremizing this action with respect to $r_H$
and $q_j$, it correctly reproduces the chemical potentials
(\ref{eq:chemp}) and the temperature (\ref{eq:invTemp}) 
of Section~\ref{sec:review}.
Secondly, the constant $3/8R$ in the definition of 
$E$ in Equation~(\ref{eq:defE}) is the Casimir energy
and it is known to exactly agree with the vacuum energy
of the large $N_c$ $\mathcal{N}=4$ super-Yang-Mills theory
defined on the manifold $\mathbb{R}\times S^3$, through
the AdS/CFT dictionary \cite{Balasubramanian:1999re}.
Apart from this important fact, this term
does not contribute to the thermodynamics of the system.
Other than this Casimir factor, the (off-shell) boundary reduced
action has precisely the same form as the on-shell Gibbs
free energy (\ref{eq:Gibbs}) with $E$
being proportional to $M$ in Equation~(\ref{eq:MinI}).%
\footnote{
  This remarkable result is also obtained in other
  systems such as the charged four dimensional black
  holes in asymptotically flat space \cite{Braden:1990hw},
  and in asymptotically AdS space \cite{Peca:1998cs}.
  Here we see that it is true even with the presence
  of the scalar fields that non-trivially couple to
  gravity and Maxwell fields.}
Finally, the mean thermal quantities can be determined by
taking derivatives of $I^*$ and evaluating them on shell
yielding
\begin{equation}\label{eq:thermalMeans}
  \langle E \rangle = E
  \;,\quad
  \langle Q_j \rangle = e_j/R^2
  \quad\text{and}\quad
  S = 2\pi \Big( \frac{r_H}{R} \Big)^3 \sqrt{H(r_H)}
  \;.
\end{equation}
The thermal quantities given above are functions of $r_H$ and
$q_j$, and one can readily verify that the thermodynamic
identity
\begin{equation}
  d \langle E \rangle = \frac{1}{\beta} d S
                        + \sum_j \mu_j \, d \langle Q_j \rangle
    \;,
\end{equation}
is satisfied.

\section{The Global Structure}\label{sec:metastability}
We now examine the global behavior of the boundary reduced action
$I^*$ in Equation~(\ref{eq:reducedAction}).
The action is a function over the parameter space
of $(r_H, q_j)$ with the fixed parameters $(\beta,\mu_j)$.
The essential features are captured by the subspace where
there is
only one non-vanishing charge (and chemical potential),
therefore, we concentrate our discussion here to this case.
In Appendix~\ref{app:metastability}, we show that the case
with general charge configurations essentially reduce to
the one charge case.
In this section, we set
\begin{equation}
  R = 1
  \;,
\end{equation}
and assume that all the length scales are measured in
units of the (asymptotic) AdS curvature radius $R$.

\subsection{The Metastability}
We first present the main result; there is a metastable
direction in the parameter space.
\begin{figure}[ht]
{
\centerline{\scalebox{0.9}{\includegraphics{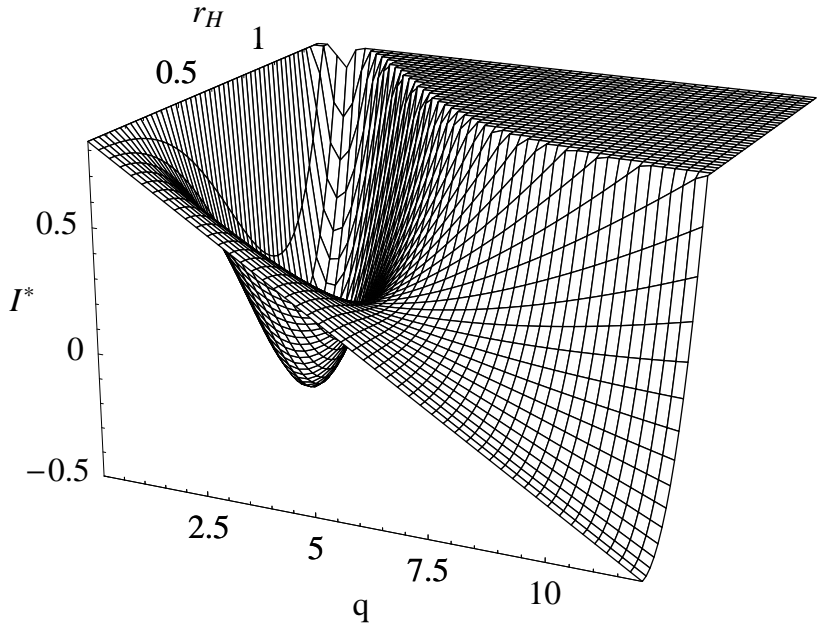}}}
\caption{\footnotesize
  The plot of the action with only one charge turned on.
  It is plotted against the variables $r_H$ and $q$,
  rather than $e$.
  The temperature and the (only one) chemical potential
  are set to $(T,\mu)=(0.45,1.05)$.
  The local minimum is located at $(r_H,q)\sim (0.996,1.23)$
  and
  the unstable saddle point is at $(r_H,q)\sim (0.359,5.38)$.
  The top flat part simply is a plot range cut-off.
}
\label{fig:metastability}
}
\end{figure}
Figure~\ref{fig:metastability} is plotted for the configuration
$(e_1,e_2,e_3)=(e,0,0)$ [or $(q_1,q_2,q_3)=(q,0,0)$ through
the relations (\ref{eq:newEQRel})].
The temperature $T:= 1/\beta$ and the only chemical potential
$\mu$ are set to $(T,\mu)=(0.45,1.05)$.
The action falls to minus infinity toward the direction
$(r_H,q)\to (0,\infty)$, therefore, there is no true
equilibrium state in this grand canonical ensemble
and the local minimum at $(r_H,q)\sim (0.996,1.23)$
is the metastable black hole.
There is an unstable saddle point at $(r_H,q)\sim (0.359,5.38)$
and it is most likely that the metastable black hole decays
by bouncing through this unstable state.
In Figure~\ref{fig:metastability}, the action appears to
go down linearly with respect to $q$.
In fact, the action has the $r_H\to 0$ limit
\begin{equation}\label{eq:tooSimple}
  I^* = \beta \Big\{
                \frac{3}{8} + (1-\mu) \, q
        \Big\}
        + \mathcal{O}(r_H^2)
        \;.
\end{equation}
This expression also shows that the {\it global} instability
sets in at $\mu = 1$. (Notice that Figure~\ref{fig:metastability}
is plotted for $\mu = 1.05 > 1$.)

The limit taken in Equation~(\ref{eq:tooSimple}) captures
the essence of the metastability but it is slightly too
simplified, so we now examine the metastable directions
in details.
For the one charge system in our current consideration,
the action $I^*$ takes the form
\begin{equation}
  I^* = \beta \bigg[ 
                \Big\{ \,
                  \frac{3}{8} + \frac{3}{2} \, r_H^2 ( 1 + r_H^2 + q )+q
                \Big\}
              -\sqrt{ q ( r_H^2 + q ) ( 1 + r_H^2 )} \; \mu
           \bigg]
           - 2 \pi \, r_H^2 \sqrt{ r_H^2 + q }
        \;.
\end{equation}
We call the first term in the square bracket, ``the mass part'',
the second term, ``the EM energy'', and the last term
in the above expression ``the entropy''.
As in the standard grand canonical ensemble, the mass part
thermodynamically competes against the EM energy and the
entropy.
We are interested in the direction in the parameter space
where the EM energy and the entropy overcome the mass
part in the action.
It is clear from the form of the action that this can
possibly happen only in the direction $q\to\infty$.
Expanding the action $I^*$ for large $q$,
we obtain
\begin{equation}\label{eq:IBar}
  I^* = \beta \Big(
                1 + \frac{3}{2} r_H^2 - \sqrt{ 1 + r_H^2 } \; \mu
                \Big) \, q
                + \mathcal{O}(q^{1/2})
        \;.
\end{equation}
The next-to-leading order is $q^{1/2}$ but
we concentrate on the leading term.
For fixed values of $\beta$, $\mu$ and $q$,
this term has the minimum
at $r_H = 0$ as long as $\mu\leq 3$, and for $\mu\geq 3$,
the minimum is located at $r_H = \sqrt{ \mu^2/9 - 1 }$.%
\footnote{
  The coefficient of the next-to-leading order is $-2\pi r_H^2$.
  Since the minimum is at $r_H=0$ as long as $\mu\leq 3$,
  this order does not modify the critical metastability line
  $\mu=1$ as in Equation~(\ref{eq:tooSimple}).
}
This means that the most likely
direction of the instability is
$(r_H,q)\to (0,\infty)$ for $1<\mu\leq 3$ and is
$(r_H,q)\to (\sqrt{ \mu^2/9 - 1 },\infty)$
for $\mu\geq 3$.
Hence, it appears that the metastable black hole decays
either by collapsing to $r_H\to 0$ with diverging charge
or into some finite horizon radius with diverging charge,
depending on the value of $\mu$.
However, we suspect that this is not the
physical behavior of the decaying black hole.
First, the physical meaning of the value $\mu=3/R$ is highly
unclear as oppose to $\mu=1/R$ which has the relation to
the BPS bound.
The second reason is that our analysis is entirely classical and
quantum corrections are not considered.
Also we have ignored the effect of the back-reaction
from the thermal gravitons.
We should expect the modifications to the action due to
those effects and should not trust the results too much
in details without good reasons.
Finally, our action can only describe the
systems satisfying the assumptions that we have
adopted along the derivation of the reduced action.
Thus it is possible that a metastable black
hole decays into some other spacetime with,
say, a different topology or a geometry that is not
spherically symmetric.
We do know that there is a metastable direction in the
parameter space but
we cannot simply conclude the decaying process of the unstable
black hole using the method adopted in this work.

To summarize, our action shows that
the metastability sets in at $\mu = 1$
as shown in Equation~(\ref{eq:tooSimple})
and a metastable
black hole decays by accumulating large amount of charge due to
the excessive value of the chemical potential.
Though we have neglected the corrections to the action due to the
quantum and back reaction effects,
we expect that those effects do not modify the qualitative
behavior of the system.
In particular, as we discussed in the introduction, we
have a good reason to expect the critical
value of the chemical potential $\mu = 1$ receives no corrections.
It is commonly assumed that an instability of a system merely implies
the existence of a new ground state
but our analysis of the theory does not suggest such picture.

\subsection{Small Black Hole, Large Black Hole and Instabilities}
We now examine the behavior of the stable and unstable
saddle points that the boundary reduced action possesses.
For the one charge configuration, our action has a
saddle point at the values of $r_H$ and $q$ that satisfy
\begin{equation}\label{eq:onShellBetaMu}
  \beta = \frac{ 2 \pi \sqrt{ r_H^2 + q } }{ 2r_H^2 + 1 + q }
  \;,\quad
  \mu = \sqrt{ \frac{ q ( 1 + r_H^2 ) }{ r_H^2 + q } }
  \;.
\end{equation}
Since we are considering the grand canonical ensemble,
these expressions should be solved for $r_H$ and $q$
in terms of given values of $\beta$ and $\mu$, and the
solutions are the thermal expectation values of the
equilibrium state.%
\footnote{
  The quantities should really be written as
  $\langle r_H \rangle$, {\it etc.}, to be consistent with
  the notations adopted at the end of the previous section.
  However we simply use the notations $r_H$, {\it etc.},
  to avoid cluttering the equations.
}
There are two relevant solutions with real and positive
horizon radii.

\begin{figure}[ht]
{
\centerline{\scalebox{1.2}{\includegraphics{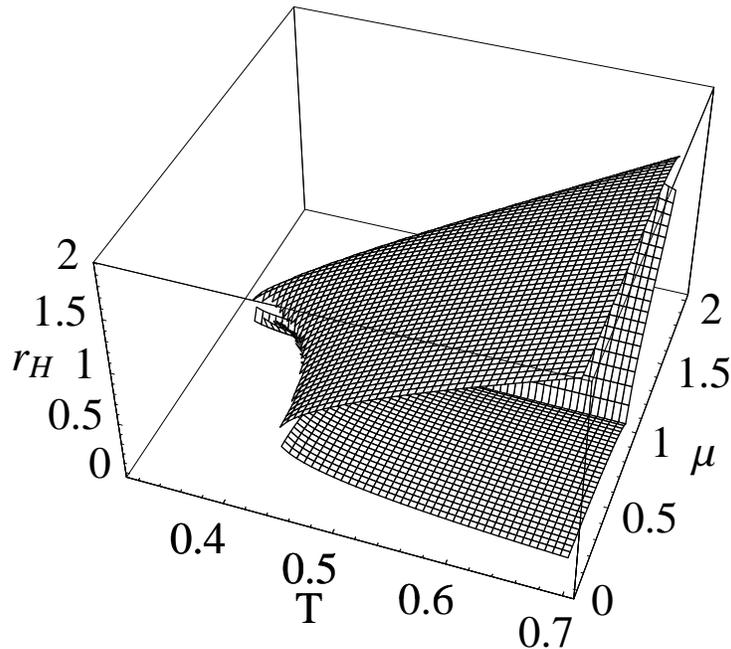}}}
\caption{\footnotesize
      The two relevant solutions, $r_H(T,\mu)$, of
      Equations~(\ref{eq:onShellBetaMu}) are plotted
      and superposed.
      The top leaf corresponds to the (locally) stable
      black hole (large black hole) and the other 
      leaf to the unstable
      black hole (small black hole).
      The lower leaf has a cusp along the line of
      $\mu = 1$ and the value of the horizon radius
      along this line is zero.
      The slight gap between the leaves is due to 
      the numerical noise.
    }
\label{fig:bigAndSmallBH}
}
\end{figure}

\begin{figure}[ht]
{
\centerline{\scalebox{1.08}{\includegraphics{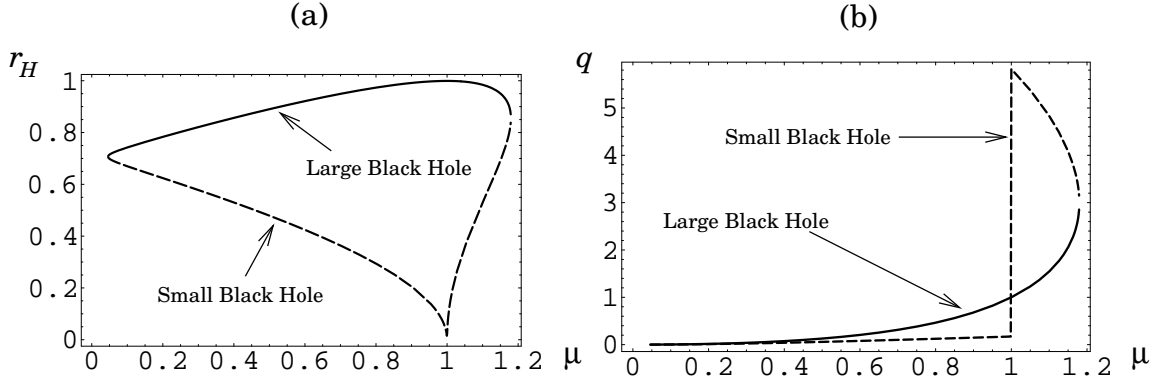}}}
\caption{\footnotesize
  Diagram (a) is plotted for the horizon radii $r_H$ of the
  large and small black holes against $\mu$ at fixed 
  temperature $T=0.45$.
  The solid line corresponds to the large black hole
  solution and the dashed line to the small black hole.
  Similar plot is given in Diagram (b) for the charges $q$
  against $\mu$ at the same temperature $T=0.45$ as the
  other diagram.
}
\label{fig:slices}
}
\end{figure}

In Figure~\ref{fig:bigAndSmallBH}, $r_H$ of those solutions
are plotted in the parameter space of $(T,\mu)$.
In the blank regions of the parameter space, there
are no saddle point solutions to the action.
The blank region that surrounds the origin of the diagram
corresponds to the part inside of the dotted line in the
left diagram of Figure~\ref{fig:holethermo}.
Also the blank region of very large $\mu$ is the local
instability part of the phase diagram that corresponds
to the region above the solid line of the left diagram 
in Figure~\ref{fig:holethermo}.
Therefore, if we project Figure~\ref{fig:bigAndSmallBH}
down to the $\mu$-$T$ plane, the boundary of the plot
is the (local) instability lines discuss in Section~\ref{sec:review}.
The top ``leaf'' of Figure~\ref{fig:bigAndSmallBH} is
the locally stable black hole solution and the other
bottom leaf is the unstable saddle point of the action.
Those two saddle points of the action are commonly called
``large black hole'' and ``small black hole'', respectively,
and we adopt this terminology here.
In Figure~\ref{fig:slices},
we also plot the slices of the functions $r_H (T,\mu)$ and
$q (T,\mu)$ against the chemical potential $\mu$
at the fixed temperature $T=0.45$.

Notice, in particular,
the behavior of the small black hole at the
line $\mu = 1$.
At this critical metastability line, the small black hole,
regardless of the temperature, shrinks to zero size
in horizon and ceases to exist.
Above the critical line, the small black hole gains
finite horizon radius again, but the charge jumps to
a bigger value
than that of the large black hole.
The large and the small black hole saddle points of
the action merge at the local instability lines
(the solid and dotted lines in the left diagram of
Figure~\ref{fig:holethermo}).

In what follows, we describe the role of the small black hole
in different phases.
In Section~\ref{sec:review}, we explained that the region
between the dotted and dashed lines in Figure~\ref{fig:holethermo}
is the spinodal phase where the thermal AdS space without
black hole is energetically preferred but the black hole
solution remains locally stable.
The action $I^*$ at the origin of the parameter space
$(r_H,q)$ corresponds to the thermal AdS space without
black hole.%
\footnote{
  Strictly speaking, the action does not describe the thermal
  AdS space because it has different topology from what is assumed
  in constructing the action.
  In fact, the origin of the action is not a saddle point,
  except when $\mu = 1$.
  However, it formally corresponds to the thermal AdS space
  with the correct equilibrium energy $3/8$ and correct black hole
  parameters of $r_H=0=q$.
}
Therefore, in the spinodal phase, the large black hole
saddle point does exist but the value of the action at
this point is larger than the value at the origin.
There is an unstable
small black hole solution between
the large black hole and the thermal AdS space
in the $(r_H,q)$ parameter space.
Thus, the large black hole in the spinodal phase is most
likely to decay into the thermal AdS space, bouncing
through the small black hole.
The behavior is opposite in
the high temperature side of the
Hawking-Page phase transition line (the dashed line
in Figure~\ref{fig:holethermo}) and below the metastability
critical line $\mu=1$.
There, the black hole solution is energetically preferred
to the thermal AdS space and the latter is most likely
to decay into the large black hole, bouncing through the
small black hole.
In the region
above the $\mu=1$ line and below the local instability
line (the solid line in Figure~\ref{fig:holethermo}),
the small black hole
is located between the large black hole and the instability
direction of the parameter space.
Therefore in this region of the phase diagram, the large
black hole is most likely to decay into the instability
direction bouncing through the small black hole.
The thermal AdS space in this phase can either directly
decay into instability direction or decay through
the large black hole.
At the local instability line present
above $\mu=1$, the large
and small black holes merge and open up the instability
direction without any barrier.

\begin{figure}[ht]
{
\centerline{\scalebox{0.9}{\includegraphics{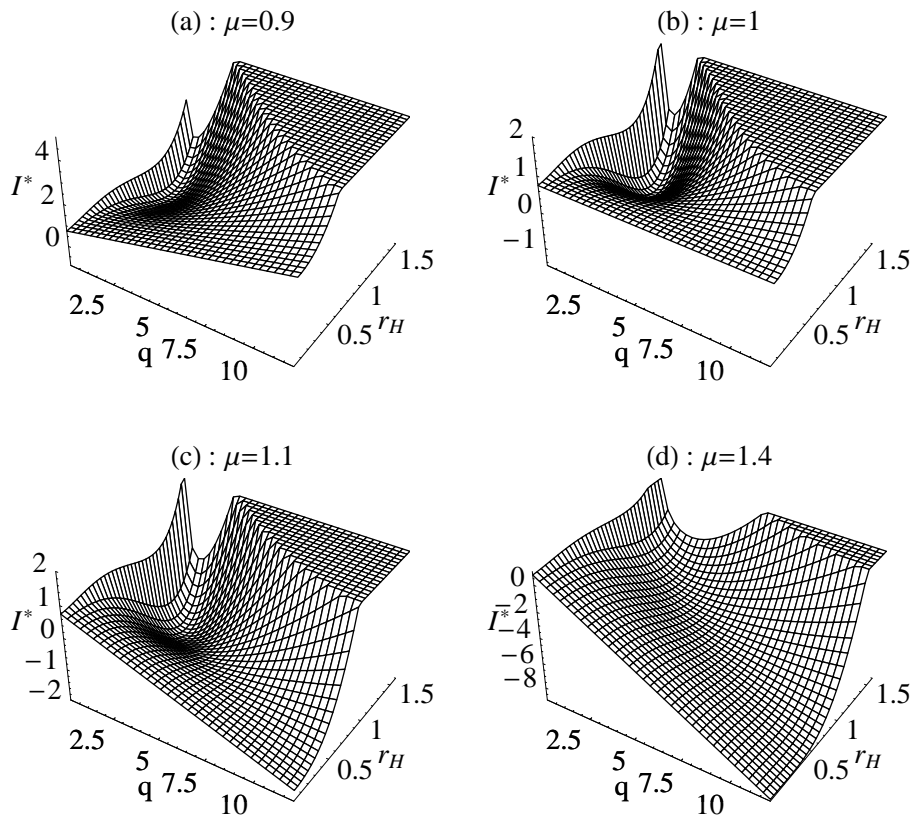}}}
\caption{\footnotesize
  The action is plotted for different values of the chemical
  potential.
  For all the diagrams, the temperature is set to $T=0.5$
  and other numerical data is shown in
  Table~\ref{table:numericalData}.  
  The top flat parts are the plot range cut-offs.
}
\label{fig:sequence}
}
\end{figure}

To illustrate the described behavior of the saddle
points,
Figure~\ref{fig:sequence} displays the sequence of the
actions with respect to rising chemical potential
and fixed temperature $T=0.5$.
The numerical data for the diagrams are summarized in
Table~\ref{table:numericalData}.
In the table, ``LBH'' and ``SBH''
stand for ``Large Black Hole'' and ``Small Black Hole'', respectively.

\begin{table}[h]
  \centerline{
  \begin{tabular}{c|c c c c}
    Diagram    & (a) & (b) & (c) & (d)  \\ \hline
    $\mu$ & $0.9$ & $1$ & $1.1$ & $1.4$	\\ 
    LBH $(r_H,q)$ & $(1.208,0.7166)$ & $(1.211,1.000)$ 
                  & $(1.205,1.415)$ & $(-,-)$	\\
    SBH $(r_H,q)$ & $(0.1087,0.1716)$ & $(0.000,-)$ 
                  & $(0.5048,6.885)$ & $(-,-)$
  \end{tabular}
  }
  \caption {\footnotesize
  	The numerical data for the plots in Figure~\ref{fig:sequence}.
  	The abbreviations ``LBH'' and ``SBH'' stand for ``Large
  	Black Hole'' and ``Small Black Hole'', respectively.
  }\label{table:numericalData}
\end{table}

We comment that 
the decay processes discussed in this section
are similar to the one analysed by
Gross {\it et al.} \cite{Gross:1982cv} where their
``Schwarzschild instanton'' corresponds to the small
black holes in our discussion.%
\footnote{
  The authors of Reference~\cite{Gross:1982cv} also discuss
  the Jeans instability.
  But as they show in the reference, this instability is due
  to the ``anti-screening'' of gravitons that induces
  negative thermal mass squared.
  Therefore this mechanism is totally different from the one
  discussed in this section.
}
The authors of Reference~\cite{Gross:1982cv} discovered
that the Schwarzschild instanton is an unstable saddle point
by examining rather general perturbation to the black hole
metric.
Here we have restricted our consideration to the static
and spherically symmetric geometries.
Moreover, we have made the very stringent form of metric ansatz
and considered very small subspace of the full parameter space.
As mentioned near the end of Section~\ref{sec:review},
reducing the parameter space usually makes it hard to find
the unstable directions.
Therefore, it is rather remarkable to find the local and global
instabilities in our narrow range of analysis.

\subsection{Below $T_0$}\label{subsec:belowTzero}
In the regions of the phase diagram discussed above, temperature
is always larger than $T_0$ at which the horizon radius
$r_H$ vanishes.
The region of the phase diagram below $T_0$ has been 
completely unknown
because the analysis was done through the on-shell action which
requires the existence of the black hole saddle point.
Now that we have the off-shell action, we can discuss this
region of the phase diagram.
We have two results in this region. 
One is that the critical
line at $\mu=1$ persists at all temperature including
below $T_0$, and the other is that the black hole saddle points
do not exist in this region.
These two facts imply that below temperature $T_0$, a thermal
AdS space decays directly into the instability direction
without going through the black hole saddle points,
provided that the chemical potential is larger than the inverse of the
curvature radius.
In fact, one can plot the action for a fixed temperature below
$T_0$ with chemical potential larger than one and finds
a similar plot to Diagram (d) of Figure~\ref{fig:sequence}.
Therefore, the critical line $\mu=1$ below $T_0$
is not the metastability
line of a black hole but it is the instability line of the
thermal AdS space.

\section{Discussions}\label{sec:discussions}

\subsection{AdS/CFT Boundary Temperature}\label{subsec:adsCftBoundaryTemp}

In Section~\ref{sec:boundaryData}, we found that the physical
boundary temperature red-shifts to zero due to the asymptotic
AdS geometry.
Therefore, in defining the boundary action, we adopted
rather {\it ad hoc} rescaled temperature given in
Equation~(\ref{eq:boundaryTemp}).
We discuss this issue more in detail and argue that the
rescaled temperature has a natural interpretation in
the sense of AdS/CFT.

\bigskip

The black hole temperature $T_H$ computed, for instance, by evaluating
the surface gravity at the horizon or by Euclidean continuation
to smooth geometry is not a physical temperature in general,
and the physical temperature depends on the
location of an observer.
Therefore, the physical temperature (sometimes called
Tolman temperature \cite{Tolman:1930})  is given by
\begin{equation}\label{eq:physicalTemp}
  T_\text{physical} = T_H V(r_B)^{-1/2}
        \;,
\end{equation}
where we have assumed a spherically symmetric black hole
geometry with radial coordinate $r$, red-shift factor
$V(r)$ and we locate the observer at $r=r_B$.

In the case of the asymptotically flat
black hole spacetime, the red-shift factor $V(r)$
can be given with the asymptotic behavior
$V\to 1$ as $r_B\to\infty$.
Thus in this case, the Tolman temperature measured at the boundary
of the space equals $T_H$ and therefore
this is commonly considered as the physical
temperature on the boundary.
The situation is different for the asymptotically AdS
black holes.
The red-shift factor in this case has the asymptotic behavior
$V\to r_B^2/R^2$ with $R$ being the curvature radius of the
asymptotic AdS geometry.
This AdS red-shift factor causes the gravitational
potential that prevents any massive object from escaping to
infinity and effectively acts as an infinite box \cite{Hawking:1982dh}.
Also we see in Equation~(\ref{eq:physicalTemp}) that the same
AdS-term red-shift the physical temperature to zero at the
boundary and thus $T_H$ is not the physical temperature of
the boundary observer.

We have mentioned in Section~\ref{sec:boundaryData} that
the AdS black hole thermodynamics in the infinite volume
has been discussed by adopting the ``renormalized''
temperature where the temperature is multiplied by the
black hole red-shift factor so that the limit $r_B\to\infty$
remains finite.
However, since
$
        T_\infty := \lim_{r_B\to\infty}T_\text{physical}V(r_B)^{1/2}
                = T_H
$,
the ``renormalized'' temperature {\it is} the temperature
$T_H$.
The original work of Hawking and Page in Reference~\cite{Hawking:1982dh}
discusses the thermodynamic properties using $T_\infty$ (or $T_H$)
and also the following works by others have frequently adopted the
same convention.
(See, for example, Reference \cite{Brill:1997mf} and the references
cited therein.)
As emphasized by many, the redefined temperature $T_\infty$
is {\it not} the physical temperature of the boundary thermodynamics
but as we stated in Section~\ref{sec:boundaryData}, there are
indications that $T_\infty$ somehow is the boundary temperature
which is consistent with the thermodynamic properties.
The reason why this is so has remained unclear.

Here, we consider this issue in the light of AdS/CFT correspondence.
Let us first recall how the conformal redundancy arises on the
boundary of AdS space, based on Reference~\cite{Witten:1998qj}.
To make our discussion concrete, let us consider the
AdS$_5$ space with the metric
\begin{equation}
	ds_\text{AdS}^2 = \frac{ R^2 }{ (1-y)^2 }
           \big[
             -(1+y)^2 dt^2 + y^{-1} dy^2
             + 4y d\Omega_3^2
           \big]
	\;,
\end{equation}
where the parameter $R$ is a length scale of the geometry
and the (dimensionless)
time coordinate $t$ is periodically identified.
The boundary is given at $y=1$ and the
metric is ill-defined for this value of $y$.
To obtain the metric that extends over to the boundary, one
can multiply the AdS metric with a function $\eta(x)^2$ where
the function $\eta(x)$ possesses a first order zero at $y=1$.
Again, for the sake of concreteness, we choose
\begin{equation}\label{eq:eta}
  \eta(x) = \frac{ e^{\omega(x)} }{ 2 } ( 1 - y )
        \;.
\end{equation}
Here, $\omega$ is an arbitrary function over the spacetime
including the boundary.
Then the boundary metric $d\tilde s^2$ is given by restricting
the extended metric to $y=1$, {\it i.e.},
\begin{equation}\label{eq:dsTilde}
  d\tilde s^2 = \big[ \eta(x)^2 d s_\text{AdS}^2 \big]_{y=1}
              = e^{2\omega} R^2 (-dt^2 + d\Omega_3^2)
              = e^{2\omega} (d\tau^2 + R^2 d\Omega_3^2)
	\;,
\end{equation}
where the function $\omega$ now is understood to be defined
over the boundary and the last equality is obtained through
the analytic continuation to the
Euclidean signature by setting $\tau = i R t$.
Observe that
the boundary metric is unique only up to
conformal transformations due to the arbitrary function
$\exp (\omega)$.
We also see that the boundary topology is $S^1\times S^3$.
{\it If} we assume the existence of a conformal field theory
on the boundary, the circumference of the $S^1$ time
circle and the length scale of $S^3$ should be respectively
interpreted as the inverse temperature and the length scale of
the space on which the CFT is supposed to live.
However, each of those quantities cannot have physical
significance due to the conformal redundancy.
A physical quantity of the field theory must be an invariant
of the redundancy and in this case, it is given by the
ratio of the $S^1$ circumference to the length scale
of $S^3$.
In other words, the temperature of the boundary field theory
should always be measured in the units of the length
scale of the spatial three sphere.

The argument made in the previous paragraph closely parallels
our problem of the black hole boundary temperature.
The ill-defined nature
of the AdS metric at the boundary is directly reflected to
the fact that
the local inverse temperature $\tilde\beta$ defined in
Equation~(\ref{eq:betaTilde}) diverges at the boundary.
Therefore, unlike the asymptotically flat case, the
thermodynamics of AdS black holes is not well-defined in
the same straightforward manner.
In order to define the black hole temperature on the AdS boundary,
one could extend the AdS metric as $d\tilde s^2$ 
in Equation~(\ref{eq:dsTilde}) and this is similar
to the definition (\ref{eq:boundaryTemp})
of the boundary black hole temperature $\beta$.
Notice that this could leave the definition of $\beta$
ambiguous by an arbitrary function.
However, the boundary metric has conformal symmetry
and thus this black hole temperature cannot be a physical one.
Therefore we realize that a possible physical black hole
temperature at the boundary must be defined as
the ratio between the circumference of $S^1$ time
circle and the length scale of $S^3$ at the boundary;
\begin{equation}
  \beta_\text{boundary} := \lim_{r_B\to\infty}
        \frac{S^1\text{-circumference at }r_B}
             {\text{length scale of }S^3\text{ at }r_B}
        = \lim_{r_B\to\infty} \frac{\tilde\beta}{r_B}
        = \beta
        \;.
\end{equation}
As we alluded in the last equality, the inverse black hole
temperature $\beta$ should really be interpreted
in this way.
[Equivalently, one may regard $\beta$ as the rescaled
temperature by an ambiguous function of the sort
shown in Equation~(\ref{eq:eta}), followed
by taking the ratio of it to the length
scale of $S^3$, which is rescaled in the same way, at the boundary.]

Now the boundary temperature thus defined may not
have any relation to the thermodynamics of the black 
hole in the bulk because it is not the Tolman
temperature of the observer at the boundary.%
\footnote{
  We should comment that in Reference \cite{Witten:1998zw},
  Witten has briefly mentioned the {\it black hole}
  temperature $\beta$ discussed here.
  However we are emphasizing that $\beta$ is not a physical
  temperature of the black hole at the boundary in a usual sense.
}
What one should be surprised here is that the boundary
action $I^*$ with the
temperature $\beta$ {\it does} consistently describe
the thermodynamics of the bulk AdS black hole.
One could speculate from this that there would be a conformal
theory on the boundary with the temperature $\beta$ and
this theory has identical thermodynamics as the bulk
black hole, {\it i.e.}, the AdS/CFT correspondence.

The definition of the
chemical potential in Equation~(\ref{eq:boundaryChemp})
also is a conformally invariant quantity.
To see this, notice that the integral of the gauge
field one-form in Equation~(\ref{eq:defMuTilde}) is a
conformal invariant.
Then the product $\tilde\mu\tilde\beta$ is invariant under
conformal transformations.
Now since we have argued that the boundary temperature $\beta$
is a conformal invariant, so must be the boundary chemical potential
$\mu$.

\subsection{More General Solution}
\label{subsec:moreGeneralSolution}

In Section~\ref{sec:constraints}, we solved the Hamiltonian
constraint equation (\ref{eq:hamilConstraint}) by adopting the
ansatz shown in Equations (\ref{eq:fXansatz}) and (\ref{eq:HAnsatz}).
Alternatively, we can take the ansatz for the functions $X_j(r)$ and 
$H(r)$ as before but do not assume the form of the function $f(r)$.
Then regard the constraint equation (\ref{eq:hamilConstraint})
as a first order differential equation of $f(r)$ and get
the solution
\begin{align}\label{eq:solutionf}
	f(r) =\,& \frac{ r^2 H(r) }
                { 3r^4 + 2 \, r^2 \sum_j q_j + q_1q_2 + q_2q_3 + q_3q_1 }
        \nonumber\\
                &\times\bigg[
                \sum_j \frac{e_j^2}{r^2+q_j}
                + r^2 \Big\{ 3 + 2 \sum_j q_j 
                        + 3 \, r^2 \Big\}
                + C
                \bigg]
	\;,
\end{align}
where $C$ is an integration constant and the length scale
is set to $R=1$.
The constant $C$ can be determined through the
condition in Equation~(\ref{eq:horizonCondition})
and we have
\begin{equation}\label{eq:constantC}
  C = - \sum_j \frac{e_j^2}{r_H^2 + q_j}
        - r_H^2 \Big\{ 3 + 2 \sum_j q_j 
                        + 3 r_H^2 \Big\}
        \;.
\end{equation}
Unlike the previous solution to the Hamiltonian constraint equation,
the parameters $e_j$ and $q_j$ are independent parameters of this
solution.
We can follow the steps taken in Section~\ref{sec:boundaryAction}
with the new function (\ref{eq:solutionf})
and obtain the boundary reduced action
\begin{align}
  I^* =& \, \beta
        \bigg[
            \frac{3}{8} + \frac{1}{2} \sum_j q_j
            + \frac{1}{2} (q_1q_2 + q_2q_3 + q_3q_1)
            + \frac{1}{2} 
            \sum_j \frac{ e_j^2 }{ r_H^2 + q_j }
            \nonumber\\
            &+ \frac{1}{2}
            \Big(
                3 r_H^4 + 2 r_H^2 \sum_j q_j + 3 r_H^2
            \Big)
            - \sum_{j=1} e_j \mu_j
        \bigg]
        - 2 \pi r_H^3 \sqrt{H(r_H)}
        \;.
\end{align}
In extremizing this action with respect to the parameters
$(r_H,e_j,q_j)$,
one finds that there are one-parameter family of extrema.
In fact, $\partial I^*/\partial r_H$ is directly proportional
to $\sum_j \partial I^*/\partial q_j$ and the equations are
not linearly independent.
We can solve the six independent equations out of the seven dependent ones
and express the solution, for example, as
\begin{subequations}\label{eq:nonunique}
\begin{align}
  \mu_j^2 &= 1 + \sum_{k \neq j} q_k + 2 r_H^2
            - \frac{2\pi r_H}{\beta}
            H_j(r_H)^{-1}\sqrt{ H(r_H) }
          \;,\\
      e_j &= \mu_j ( r_H^2 + q_j )
      \;.
\end{align}
\end{subequations}
Using a part of equations of motion, we can determine the
Lagrange multipliers $b(r)$ and $A_{j\,\tau}(r)$ as
\begin{equation}
  b = \frac{\beta}{2\pi} f^{1/2} H^{-1/3}
  \;,\quad\text{and}\quad
  A_{j\,\tau} = -i \, \frac{\beta}{2\pi}
                \Big(
                  \frac{e_j}{r_H^2 + q_j} - \frac{e_j}{r^2 + q_j}
                \Big)
      \;.
\end{equation}
We thus have forms of all the fields and it is
checked that they, together with the the relations (\ref{eq:nonunique}),
satisfy all the equations of motion including the Maxwell and
scalar field equations with a free parameter, say, $\beta$.%
\footnote{\label{foot:analyticCont}
  We note that the gauge field shown above differs from
  Equation~(\ref{eq:lorentzianA}) by
  the factor of $-i\beta/2\pi$ which comes from the analytic
  continuation as shown in Equation~(\ref{eq:continuationA}).
}
Furthermore, we have computed the surface gravity at the horizon
(with the same analytic continuation as the one mentioned in
footnote \ref{foot:analyticCont})
and consistently obtained
\begin{equation}
  \kappa = 2 \pi / \beta
  \;.
\end{equation}
Also the thermodynamic identity can be shown to be satisfied,
just as we have done at the end of Section~\ref{sec:boundaryAction}.
Parenthetically, the solution satisfies the smoothness condition
(\ref{eq:cigarCondition}) as we imposed in obtaining the reduced
action.

The asymptotic behavior of the function $f(r)$ is
\begin{equation}
  f(r) = r^2 + \Big( 1 + \sum_j q_j \Big)
         + \mathcal{O}\Big( \frac{1}{r^2} \Big)
       \;.
\end{equation}
Up to this order, this agrees with the asymptotic behavior of
the solution obtained in Section~\ref{sec:constraints} and in
particular, the first term is
consistent with the asymptotic condition (\ref{eq:asymptAdSCond}).
One can check that the function $f(r)$ takes the form in
Equation~(\ref{eq:fXansatz}) if and only if
the relations (\ref{eq:newEQRel}) hold.
So while new $f(r)$ is parametrized by the independent
parameters $(r_H,e_j,q_j)$, the previous one
is recovered by restricting the new function to the
subset of the parameters by the relation (\ref{eq:newEQRel}).
As we have seen in Section~\ref{sec:boundaryAction}, this
relation reproduces the solution discussed in Section~\ref{sec:review}.
Therefore, we have obtained the more general solution that
include the previous solution as a special case.

What is perhaps odd about this solution
is that the grand canonical thermodynamic data,
$(\beta,\mu_j)$, cannot uniquely specify the properties of the
black hole.
Stated differently, the black hole can be uniquely
characterized by the parameters, $(E , e_j, q_1)$,
where $E$ is the thermodynamic mean 
energy derived from the action and
we have chosen $q_1$ to be the free parameter of the degenerate
solution (\ref{eq:nonunique}).
Thus, the degeneracy seems to indicate a scalar hair
of the black hole.
But this conclusion sharply contradicts with the no-hair
theorem of Sudarsky and Gonzalez \cite{Sudarsky:2002mk}.
They have shown that there cannot be scalar hairs for an
asymptotically AdS black hole where the scalar potential
approaches the global minimum near the boundary.
Our scalar potential $V$ in (\ref{eq:potential}) with 
the solutions $X_j$ in (\ref{eq:fXansatz}) actually has
such asymptotic behavior.
How can this be?
One possibility is that the degenerate solution presented
in this subsection is equivalent to the known solution
through some reparametrization (we have not been able to
find such reparametrization).
Another possibility is that the no-hair theorem is somehow
evaded in our theory, for example, due to the coupling of
the scalar to the Abelian gauge fields
(such coupling was not considered in
Reference\cite{Sudarsky:2002mk}).
Clarifying this issue
requires further considerations separate from
the topic of this work.

\subsection{On Finite Counterterm}\label{subsec:finiteCounterterm}

In Section~\ref{sec:boundaryAction}, we added the term called
``finite counterterm'' as shown in
Equation~(\ref{eq:finiteCounterterm}).
The original motivation for including the finite
counterterm is as follows.
Buchel and Pando Zayas \cite{Buchel:2003re} observed that
when the mass of the $R$-charged AdS black hole was computed
using the standard counterterms, an unexpected nonlinear
term with respect to the charge appeared.
(The charge referred here is the parameter $q$ and
not $e$.)
This nonlinear behavior of the charge in the mass can
contradict the BPS-expected inequality between the mass and the charge,
and more importantly, the thermodynamic identity is violated.
By requiring the identity to hold, they added a term that
consequently eliminated the nonlinear term.
Then Liu and Sabra pointed out in Reference~\cite{Liu:2004it}
that the inclusion of the counterterm associated with
the presence of the scalar fields is possible.
They showed that the mass computed with the finite counterterm
(\ref{eq:finiteCounterterm}) reproduced the result obtained
by Buchel and Pando Zayas.

The counterterms that arise from the presence of matter
fields are further studied by Batrachenko {\it et al.}
in Reference~\cite{Batrachenko:2004fd} utilizing the
Hamilton-Jacobi method.
In this approach, the counterterms are treated uniformly
throughout different spacetime dimensions and in particular,
they have found that the scalar field counterterms are
divergent, finite and vanishing for the spacetime dimensions
four, five and higher, respectively.
Note that the inclusion of the scalar counterterm in four
dimension is necessary to render the action finite.
Thus, it appears natural to include the scalar counterterm
for the case of five dimensions but it still requires
physical interpretation.

Liu and Sabra in Reference~\cite{Liu:2004it} mentioned
that the option to include (or not to include) the finite
counterterm in five dimension can be interpreted as the
counter part to the choice of a renormalization scheme
in a field theory.
Since the inclusion of the finite counterterm results in
the mass that has the BPS-like mass-charge relation,
they argued that the counterterm corresponds to the
renormalization scheme that preserves the supersymmetry
of the boundary field theory.
If this interpretation is valid, the physical observables
should not be affected by the finite counterterms just as
the choice of the renormalization scheme does not affect
the physical observables of a field theory.

In our work, we have derived the action that describes the
thermodynamics of the black hole observed at the boundary.
The AdS/CFT conjecture postulates that the exponential
of (minus) the action {\it is} the partition function
of the dual field theory at strong coupling.
Then the thermal mean values derived from the action,
such as $\langle E \rangle$ or $\langle Q_j \rangle$
of Equation~(\ref{eq:thermalMeans}) must be the physical
observables of the thermal field theory.
However, we see that the addition of the finite counterterm
does affect the thermal mean values.
It is most clearly seen in the charges
\begin{equation}
  \langle Q_j \rangle = e_j =
  \sqrt{q_j (r_H^2 + q_j)\bigg\{ 1+r_H^2\prod_{k\neq j}H_k(r_H) \bigg\}}
  \;.
\end{equation}
This expression is derived from the metric ansatz and the Hamiltonian
constraint equation, so the form is independent of the counterterms.
But the numerical value does change by the presence
and absence of the finite counterterm because the location
of the extremum is affected.
Similar argument applies to the other mean values.
Therefore, the observables of the theory do seem to depend
on the finite counterterm.

The dependence may imply that the finite counterterm has
different physical interpretation.
Another possibility is that the contributions to the partition
function from other saddle points may make the physical
observable independent of the finite counterterm.
In any case,
the observation made here urges us to investigate the physical
interpretation of the finite counterterm and better understanding
is desired.

\subsection{Comparison to the Field Theory Analysis}\label{subsec:comparison}
\begin{figure}[ht]
{
\centerline{\scalebox{.75}{\includegraphics{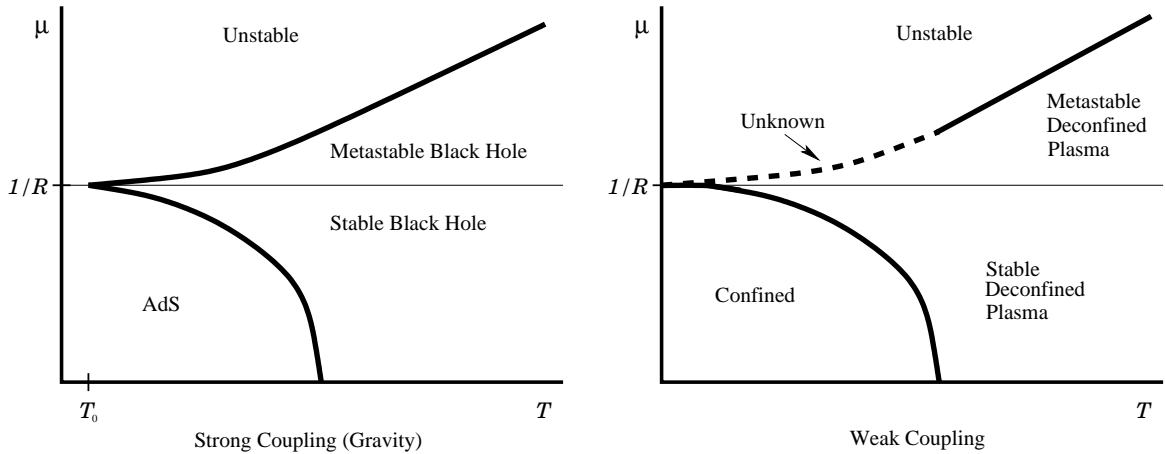}}}
\smallskip
\caption{\footnotesize 
  The schematic phase diagram of the $R$-charged black
  hole is shown on the left-hand side and on the right-hand side,
  that of
  the field theory which is supposed to be the
  AdS/CFT dual is shown.
  According to the AdS/CFT correspondence, the black hole
  system is identical to the strongly coupled region of
  the field theory, as indicated in the diagram.
  For both figures, $\mu$ is the largest of the three
  chemical potentials.
  In the strong coupling diagram,
  $T_0$ is the temperature at which
  the black hole radius reaches to zero.
  Its value depends on the pattern of chemical potentials
  and equals
  $(\pi R)^{-1}$ for a single non-zero charge, while
  $T_0=0$ for three equal charges.
  On the other hand, the nature of the instability in the
  low temperature regime of the weakly coupled field theory
  is yet to be explored and is indicated by the dashed line
  in the diagram.
}
\label{fig:comparison}
}
\end{figure}

In this work, we have found an additional structure in the phase
diagram of $R$-charged black hole system.
The phase diagram in Figure~\ref{fig:holethermo} is now augmented
by the line $\mu=1/R$ that separates the stable and metastable
black hole states for $T>T_0$ and below $T_0$, the same line
is the instability critical line of the thermal AdS space.
As mentioned in the introduction, the existence of the metastable
black hole has been anticipated from the AdS/CFT-conjectured
field theory analysis in Reference~\cite{Yamada:2006rx}.
The current status of the phase diagrams in the gravity and
field theory sides is schematically summarized in
Figure~\ref{fig:comparison}.
We see the agreement in the general structure of those phase
diagrams and the additional match in the metastability critical line
$\mu=1/R$ is the improvement made in this work.

In Section~\ref{subsec:belowTzero} we saw that
in the region of the phase diagram below $T_0$, a thermal
AdS space decays directly into unstable direction without
going through the black hole saddle points.
The corresponding field theory analysis has not been carried out
at the low temperature.
However, given the result of the gravity analysis, we have
some expectations for the behavior of the field theory.
First we note that the ``confinement/deconfinement''
phase transition
line is obtained in the zero coupling limit of the field theory.%
\footnote{
  The $\mathcal{N}=4$ super-Yang-Mills theory does not confine.
  The terminology ``confinement'' is used in accordance with the
  behavior of an order parameter, the Polyakov loop.
  See Reference~\cite{Yamada:2006rx} for more explanations.
}
The phase transition line and the $\mu = 1$ line merge exactly
at $T=0$.
The zero coupling limit is taken so that the resulting free
theory is continuously connected to the theory with finite
coupling constant.%
\footnote{
  Technically, the zero coupling limit is taken while keeping
  the Gauss law constraint.
}
Thus we expect the general structure of the phase transition
line to remain the same at a finite coupling, but it is likely
that the details would change.
For instance, it is possible that the phase transition line
at the finite coupling is modified so that it merges with
the instability line at some finite temperature.
This behavior could be similar to the one observed in the
gravity analysis.
To verify this expectation, one needs to find the
confinement/deconfinement
phase transition line at weak but finite coupling with
nonzero chemical potentials and also the instability line at
the low temperature region must be determined.
We believe that the analysis, especially
in the low temperature limit, is
doable and the work is in progress.

In the field theory analysis of
Reference~\cite{Yamada:2006rx}, it has been found that
the decay of a metastable state
is most likely to occur by one of the
eigenvalues of the scalar field splitting up from the rest of
the eigenvalues in the field space and indefinitely growing.
Our analysis in this paper does not necessarily suggest
this mechanism.
If we attempt to interpret this phenomenon in terms of
string theory, it would correspond to a D3-brane separating
from a stack of rotating D3-branes due to the high angular velocity of
the D3-branes in their transverse directions.
Therefore, we could also detect the metastability by examining
the behavior of a probe D3-brane immersed in the background
of the $R$-charged black hole or of the rotating D3-branes
(before taking the decouping limit).
This scenario is currently being investigated.

Both gravitational and field theory analysis show that two theories
cease to exist above the critical value of the chemical potential
where the local instability sets in.
In the AdS/CFT context, this means that the theory at strong
and weak coupling regimes both become unstable at a large value
of the chemical potential.
One might naively interpolate the results and speculate that
the theory in any coupling regime is unstable beyond a certain
value of the chemical potential.
Though we think that this is the most likely scenario,
we still cannot completely exclude the possibility that the theory
develops a new ground state beyond the instability line
and defines a Higgs phase.
As mentioned in Section~\ref{sec:metastability}, we did not
take into account of the back reaction from the thermal radiation
and also the quantum corrections are neglected.
(Needless to say, stringy effects have not been considered
which we expect to be important when $r_H$ becomes comparable
to the length scale of string theory.)
On the field theory side, the analysis has been carried out
at one-loop level and higher loop with finite $N_c$
effects are still unknown.
They may or may not change the global behavior of the potential
in the theory.

\bigskip
\bigskip

\section*{Acknowledgments}

I would like to thank Gerold Betschart, Shmuel Elitzur,
Barak Kol and Eliezer Rabinovici
for discussions and valuable comments.
I would also like to thank S. Elitzur for reading through the
manuscript.
This work was supported by the Golda Meir Post-Doctoral fellowship.

\bigskip
\bigskip

\appendix

\section{Metastability in General 
Charge Configuration}\label{app:metastability}
In the main text, the metastability of the $R$-charged black hole
is examined in the subspace of the full parameter space
with the only one nonzero charge.
This appendix augments the argument that the metastable black hole
system with a general charge configuration tends to have the
unstable direction into the single charge configuration.

\begin{figure}[ht]
{
\centerline{\scalebox{0.7}{\includegraphics{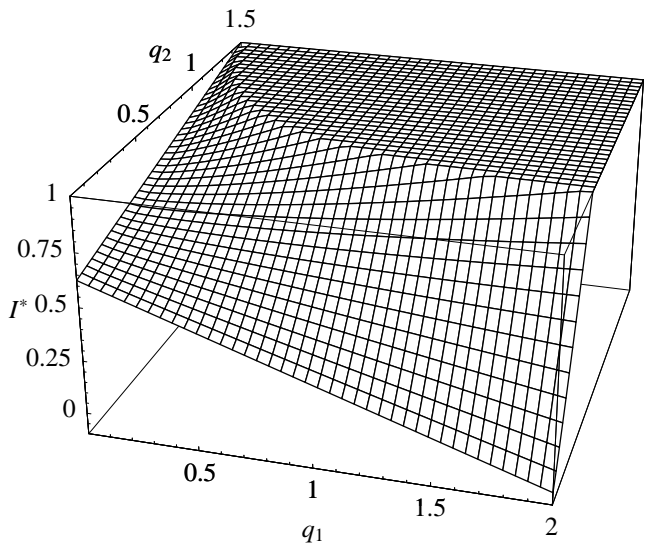}}}
\caption{\footnotesize
      In this plot of the action, the configuration with
      $q_3 \equiv 0$ is chosen.
      The temperature, the horizon radius and the chemical
      potentials
      are set to $T = 0.6$, $r_H = 0.01$ and
      $(\mu_1,\mu_2)=(1.2,0.85)$, respectively.
      The diagram shows that an instability exists only in the
      direction of the large charge whose corresponding chemical
      potential exceeds the critical value $1$.
      It also exhibits that the direction along $q_1 = q_2$
      is stable.
      The top flat part is a plot range cut-off.
    }
\label{fig:appendix}
}
\end{figure}

As in the single charge case of the main text, a possible
instability occurs only when one of the charges diverges.
This is because the $r_0$ factor in $E$ of Equation~(\ref{eq:reducedAction})
overwhelms the rest of the terms
when more than one charges become large.
Therefore, without loss of generality, we expand the action
with respect to large $q_1$ to obtain
\begin{equation}
  I^* = \beta \Big(
                1 + \frac{3}{2} x_1^2 - \sqrt{ 1 + x_1^2 } \; \mu_1
                \Big) \, q_1
                + \mathcal{O}(q_1^{1/2})
        \;,
\end{equation}
where we have defined the parameter
$x_1 := r_H^2 H_2(r_H) H_3(r_H)$.
This is exactly the same form as in Equation~(\ref{eq:IBar})
and therefore, for fixed values of $\beta$, $\mu_1$ and $q_1$,
the leading term has the minimum
at $x_1 = 0$ as long as $\mu_1\leq 3$, and for $\mu_1\geq 3$,
the minimum is located at $x_1 = \sqrt{ \mu_1^2/9 - 1 }$.
This implies that as long as $1<\mu_1\leq 3$, the unstable
direction is $(r_H,q_1,q_2,q_3)\to (0,\infty,0,0)$ and for
$\mu_1\geq 3$, the parameters $r_H$, $q_2$ and $q_3$ may
be nonzero but none of them may diverge.
One of the consequences of this consideration is that
the constrained system with
$q_1 \equiv q_2\,,\; q_3\equiv 0$ or $q_1 \equiv q_2 \equiv q_3$
does not have the global instability (and 
as mentioned near the end of
Section~\ref{sec:review}, these configurations do not have the
local instabilities as well).
Those constraints reduce the dimension of the parameter space
and the unstable directions are closed.
The discussion made in this paragraph is illustrated in
Figure~\ref{fig:appendix} for the case with the two nonzero
charges.
Observe that the instability direction opens up for the 
large charge whose conjugate chemical potential is larger
than one.
In this plot, it is the direction of large $q_1$ and not
$q_2$.
Also note that the action does not have instability along
the line $q_1 = q_2$.

The result obtained in this appendix confirms the claim
made in the main text; the metastability in
the case with general charge configurations
essentially boils down to the system with one charge.

\pagebreak


\end{document}